# Multifunctional Oxides for Topological Magnetic Textures by Design

Z. S. Lim[1,2], A. Ariando[1,2,3,4]*

[1]Department of Physics, National University of Singapore, Singapore 117542, Singapore
[2]NUSNNI-NanoCore, National University of Singapore, Singapore 117411, Singapore
[3]Centre for Advanced 2D Materials and Graphene Research Centre, National University of Singapore, Singapore 117551
[4]NUS Graduate School for Integrative Sciences and Engineering, National University of Singapore, Singapore 117456, Singapore

E-mail: ariando@nus.edu.sg

**Abstract**

Several challenges in designing an operational Skyrmion racetrack memory are well-known. Among those challenges, a few contradictions can be identified if researchers were to rely only on metallic materials. Hence, expanding the exploration on Skyrmion Physics into oxide materials is essential to bridge the contradicting gap. In this topical review, we first briefly revise the theories and criteria involved in stabilizing and manipulating Skymions, followed by studying the behaviors of dipolar-stabilized magnetic bubbles. Next, we explore the properties of multiferroic Skyrmions with magnetoelectric coupling, which can only be stabilized in $Cu_2OSeO_3$ thus far, as well as the rare bulk Néel-type Skyrmions in some polar materials. As an interlude section, we review the theory of Anomalous (AHE) and Topological Hall Effect (THE), before going through the recent progress of THE in oxide thin films. The debate about an alternative interpretation is also discussed. Finally, this review ends with future outlooks about the promising strategies of using interfacial charge-transfer and (111)-orientation of perovskites to benefit the field of Skyrmion research.

Keywords: Oxide Skyrmions, Topological Hall Effect, Thin film heterostructures

## 1. Introductions

### 1.1 Essential Physics of Skyrmions

The concept of topological defects as soliton solutions to a continuous field is present in a broad scope of natural Physics, including tornadoes, whirlpools, Abrikosov vortices in type-II superconductors, liquid crystals, etc. that involve a non-linear field equation. Skyrme first proposed such a quasiparticle theory to predict several properties of nucleons as excitation in pion fields[1]; while Skyrmions in magnetism is physically real and potentially useful in spintronics. Magnetic Skyrmion is a quasiparticle of collective magnetic moment excitations with a topological winding number, Q that fully wraps around a Bloch-sphere for integer number of times (Fig. 1a-c):

$$Q = \frac{1}{4\pi}\int \hat{\boldsymbol{m}} \cdot \left(\frac{\partial \hat{\boldsymbol{m}}}{\partial \mathrm{x}} \times \frac{\partial \hat{\boldsymbol{m}}}{\partial \mathrm{y}}\right) \mathrm{dxdy} = \pm\mathrm{integer} \qquad (1)$$

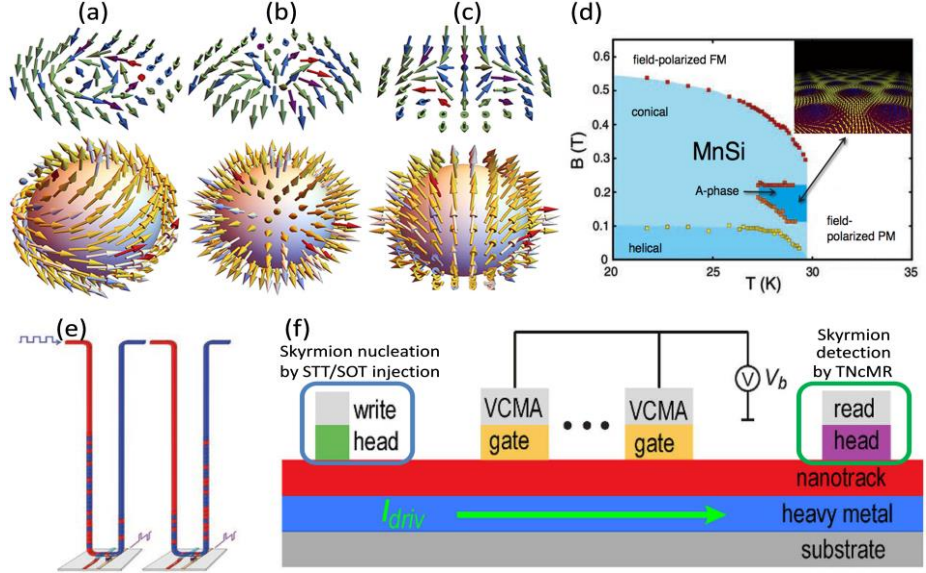

**Figure 1.** Skyrmions with topological parameters ($Q$, $Q_v$, $Q_h$): (a) (1,1,π/2), (b) (1,1,0) and (c) (-1,-1,0). The top panel shows Skymions in 2D while the bottom panel shows their respective stereographic projections into 3D Bloch spheres. (d) A typical magnetic phase diagram of bulk MnSi. (e) Vertical domain-wall racetrack memory array proposed by Stuart Parkin. The domain wall information storage bits can be replaced by Skyrmions. (f) An illustration of a Skyrmion racetrack memory highlighting the essential parts: (left to right) nucleation at WRITE head by localized current injection or dissipation-less electric field, VCMA top-gates, Skyrmion transport racetrack with the heavy metal layer providing STT or SOT, and READ head for Skyrmion detection via TNcMR. *Reprinted (adapted) with permission from Ref.* [2-5]. *Copyright (2018) Kovalev and Sandhoefner, (2008-2009) Science AAAS and (2016) Nature Publishing Group.*

In a typical magnetic system, the relevant atomic-scale energy landscape is described by an effective Hamiltonian is:

$$\widetilde{H} = -J_{ex}\sum_{ij}\widehat{\boldsymbol{m}}_i \cdot \widehat{\boldsymbol{m}}_j - K_u\sum_i m_{z,i}^2 - \sum_{ij}\boldsymbol{D}_{ij}\cdot(\widehat{\boldsymbol{m}}_i \times \widehat{\boldsymbol{m}}_j) - \mu_o \boldsymbol{H}\cdot\sum_i\widehat{\boldsymbol{m}}_i \qquad (2)$$

where the 1st to 4th terms are isotropic exchange, uniaxial anisotropy, Dzyaloshinskii-Moriya interaction (DMI) and Zeeman energy. There are two types of spin-orbit coupling (SOC): the Rashba-type $\alpha_R(\boldsymbol{k}\times\hat{\boldsymbol{z}})\cdot\hat{\boldsymbol{\sigma}}$ and the Dresselhaus-type $\alpha_D[k_x(k_y^2 - k_z^2)\sigma_x + k_y(k_z^2 - k_x^2)\sigma_y + k_z(k_x^2 - k_y^2)\sigma_z]$ that emerge from broken interface and bulk inversion symmetry respectively, where k, σ and $z$ are the electron's momentum, spin and the sharp interface normal vector. Correspondingly, there are two types of DMI that would stabilize the Néel and Bloch types Skyrmions[6-8]. The DMI term in Eqn. 2 can be rewritten into a linear combination of $D\sin(\gamma)\,\widehat{\boldsymbol{m}}\cdot[(\hat{\boldsymbol{z}}\times\boldsymbol{\nabla})\times\widehat{\boldsymbol{m}}] + D\cos(\gamma)\,\widehat{\boldsymbol{m}}\cdot(\boldsymbol{\nabla}\times\widehat{\boldsymbol{m}})$, where $\tan(\gamma) = \frac{D_\parallel}{D_\perp}$ is a ratio between the Rashba-type and Dresselhaus-type DMI, emerging in crystals with $C_{nv}$ and $D_n$ symmetries respectively[9, 10]. Hence, the Bloch-type Skyrmions can be found in materials with non-centrosymmetric or hexagonal crystal structures[3, 11, 12]; while the Néel-type ones exist at sharp interfaces between FM and paramagnetic (PM) heavy metal without specific requirement on crystal structures[13-16]. From Eqn. 1, Skyrmions can be nucleated, usually assisted by application of a small magnetic field, when a criterion is fulfilled[17]:

$$\frac{D}{\sqrt{J_{ex}K_u}} > \frac{4}{\pi} \qquad (3)$$

This is also coined as the effective anisotropy model[10]. The resulted Skyrmion radius is $r_{sk} = \sqrt{J_{ex}}/\sqrt{2K_u(1 - D/D_c)}$ where $D_c = 4\sqrt{J_{ex}K_u}/\pi$. From Landau-Ginsburg framework, 3 helicoids merge to form a Skyrmion-lattice (SkL) assisted by an out-of-plane magnetic field[18, 19]. This hence results in a typical Skyrmion phase diagram: at zero/low magnetic field, helical or cycloidal domains exist in labyrinth-like stripes; SkL phase exists at intermediate field; at large field, moments are saturated into collinear ferromagnetic (FM) phase (Fig. 1d). A complete theoretical categorization of |Q|=1 Skyrmions can be done according to three topological parameters: the winding number, vorticity and helicity ($Q$, $Q_v$, $Q_h$), which results in 16 configurations in total including antiskyrmions[20, 21]. Other topologies such as Merons ($Q=\pm 0.5$), Skyrmionium ($Q=0$) and bi-Skyrmion ($Q=\pm 2$) are also possible, though rare.

*1.2 Design Criteria for Skyrmion Memory*

From a practical application viewpoint, Stuart Parkin first proposed the concept of domain-wall (DW) racetrack memory on year 2008 (Fig. 1e), where the DW movements are driven by current via spin-transfer torque (STT)[4, 22, 23]. Subsequent research motivation in this field placed emphasis on reducing the threshold current density ($j_c$) for initiating the DW motion, below which the DWs will be pinned by impurity/defects. Typical $j_c$ with STT on soft collinear FMs such as permalloys is $10^{11}$–$10^{12}$ A/m$^2$. However, Bloch-type Skyrmions in MnSi and FeGe single crystals[24-26] have demonstrated an astounding low $j_c$ of merely ~$10^6$ A/m$^2$, suggesting that replacing FM DWs with Skyrmions as information storage bits is an ideal strategy for building future energy-saving racetrack memories. In the route of achieving an operational Skyrmion racetrack memory[5, 27], several challenging aspects need to be overcome as listed below:

(1) Individual Skyrmions should be controllably nucleated at the WRITE-head by providing a perturbation to a local FM energy landscape. Thus far, the two successful schemes are: (i) application of an electric field via a sharp, antiferromagnetic (AF) Cr scanning tunnelling microscope (STM) probe tip[28, 29], and (ii) injection of a nanosecond pulsed current via a nano-patterned injector electrode[30]. For scheme (i), a near bi-stable energy landscape between collinear FM and Skyrmion is first achieved by applying a small magnetic field, and subsequent injection of spin-polarized tunnelling electrons into the sample would excite across the energy barrier for Skyrmion nucleation due to an STT effect. Changing the spin-polarized Cr tip into an un-polarized W tip also demonstrated the pure electric field switching effect[29]. While for scheme (ii), a charge current injection in the x-direction ($J_{c,x}$) of a heavy metal (e.g. Pt) would produce a spin current in the z-direction, carrying y-direction spin ($J_{s,z}^y$) by Spin Hall Effect (SHE). It then exerts a Spin-orbit Torque (SOT) onto the adjacent FM with perpendicular magnetic anisotropy (PMA) ($M_z$), assisting to nucleate Skyrmions with opposite-moment core with respect to the surrounding magnetization[31, 32]. To enhance the spin accumulation density by SHE, a delicately-patterned sharp current injector or constriction is desirable, so that Skyrmions always nucleate at a predictable position.

To minimize energy consumption by achieving zero-current Skyrmion nucleation, an intrinsic material property – magnetoelectric coupling[12, 33] allowing electric field control of magnetism in multiferroic Skyrmions such as $Cu_2OSeO_3$ – is valuable but remains very rare. Besides, two more possible Skyrmion tuning mechanisms by an electric field gating have gained attention. In the voltage-controlled magnetocrystalline anisostropy (VCMA) design, the electric field is expected to tune the carrier density significantly (typical at a ferromagnet/oxide interface) and induce a change in magnetic anisotropy locally in the nanotrack. Maruyama discussed that a negative voltage may lift the degeneracy of 3d orbitals – rising the energy of $d_{3z^2-r^2}$ ($L_z$=0) states, populate more electrons into $d_{x^2-y^2}$ ($L_z$=2) states and hence increases the PMA[34]. This mechanism can be used to control the Skyrmions' motion by creating a trapping/de-trapping potential beneath the top-gate (Fig. 1f)[5, 27], or function as the WRITE-head to create/annihilate Skyrmions[35]. A magnetic-field-free scenario of such operation is possible by providing an exchange bias field from an adjacent antiferromagnet. On the other hand, a tight-binding theoretical calculation expected a change in intra-orbital hopping amplitude and Rashba-type SOC[36], but is nevertheless a very weak (2$^{nd}$-order perturbation) effect if magnetoelectric coupling is absent. In short, we can see that all strategies involved tunings around fulfilling the Eqn. 3 criterion.

(2) Skyrmions should be driven by low current to drift along the racetrack[24-26]. From a simple power dissipation ($I^2R$) concept, reducing drive-current is the key to develop low energy-consumption memory, with contemporary criterion in the atto-joule ($10^{-18}$ J) range. The Thiele-modified Landau-Lifshitz-Gilbert (LLG) equation is usually employed in simulating Skyrmions' motion[37]:

$$\boldsymbol{G} \times (\boldsymbol{v}_e - \boldsymbol{v}_{sk}) + \boldsymbol{D} \cdot (\beta \boldsymbol{v}_e - \alpha \boldsymbol{v}_{sk}) + F_{pin} = 0 \qquad (4)$$

Where the 1$^{st}$ and 2$^{nd}$ terms are the Magnus and spin-transfer forces respectively, with the gyromagnetic vector $\boldsymbol{G} = 4\pi Q \hat{\boldsymbol{z}}$, and the dissipative tensor $\boldsymbol{D} = \begin{bmatrix} \int (\partial_i \hat{\boldsymbol{m}} \cdot \partial_j \hat{\boldsymbol{m}}) \mathrm{d}^2 r & 0 \\ 0 & \int (\partial_i \hat{\boldsymbol{m}} \cdot \partial_j \hat{\boldsymbol{m}}) \mathrm{d}^2 r \end{bmatrix}$, $\boldsymbol{v}_e$ and $\boldsymbol{v}_{sk}$ are the drift velocities of conduction electrons and Skyrmions, and $\alpha, \beta$ are the adiabatic and non-adiabatic Gilbert damping parameters. $F_{pin}$ in the 3$^{rd}$ term is the pinning force from impurity/defect or control of anisotropy. For simplicity, the SOT term[38, 39] in Eqn. 4 is dropped. In a typical micromagnetic simulation involving Eqn. 4, the ultralow $j_c$ was understood as Skyrmions' ability to evade impurity/defect pinning due to their topological protection[40]. A pulsed current injection is the common engineering strategy to reduce Joule heating, by using a low duty cycle = pulse width/interval[30, 32, 41, 42]. Conversely, a DC current would always cause heat accumulation and continuous rise of device temperature with time, resulting in uncontrollable thermal-induced nucleation, size change or deletion of magnetic Skyrmions. For example, in the thermally-driven Skyrmion nucleation scheme, the threshold nucleation current can be seen varying inversely with duty cycle[41]. Pulsed-current injection also facilitates quantification of Skyrmions'/domain walls' velocities by static position imaging[43, 44].

(3) The Skyrmions' motion across the racetrack has to be straight to avoid annihilation at the sidewall. From Eqn. 4, the Magnus force (1$^{st}$ term) is expected to produce a transverse Skyrmion velocity component ($v_{sk,\perp}$) named as the Skyrmion Hall Effect (SkHE)[43, 44]. Again a pulsed current is beneficial to avoid annihilation at sidewalls, since Skyrmions would be repelled by sidewalls if the Magnus force is below a certain limit. Ferrimagnetic Skyrmions such as GdFeCo have also been shown to mitigate the SkHE angle[45]. To totally eliminate this SkHE, search for antiferromagnetic Skyrmions[46, 47] and Skyrmioniums[48] (both have Q=0) have been proposed but remained elusive; until only recently a synthetic antiferromagnetic (SAF) multilayer Skyrmion structure bound by Ruderman–Kittel–Kasuya–Yosida (RKKY) interaction[15] was found to be stable.

(4) Skyrmion detection at the READ-head should be via Tunnelling Non-collinear Magnetoresistance (TNcMR), which is a tool compatible with practical device miniaturization. A tunnel barrier should be inserted between the Skyrmion racetrack and a PM metal. Due to difference in density-of-states (DOS) between collinear FM and Skyrmion textures, a change in tunnelling resistance can be detected[29, 49].

*1.3 Why oxides? Motivations and Opportunities*

- Oxides have varieties of physical interactions – super-exchange, strong SOC, strain effect on crystal field splitting, and tunable Mott-Hubbard interaction, etc.
- Varieties of oxides are antiferromagnetic, bringing benefit for aspect (3).
- Oxides also have lower and tunable carrier density than typical metals, and more susceptible to electric field gating for aspect (1). The ferroelectric or multiferroic magnetoelectric coupling can be in found in various oxides, allowing non-volatile, zero-current tuning of magnetic properties.
- Oxides may have complex non-centrosymmetric crystal structures that contribute Dresselhaus DMI.
- Interfacial or Rashba-type DMI could be controlled via termination engineering in perovskite oxides, such as polar/nonpolar AO & $BO_2$ in (001), or polar $AO_3$ and B in (111) orientations.
- Perovskite oxide thin film interfaces can be fabricated with high quality for an interesting search of composite Néel and Bloch types Skyrmions.
- Skyrmions hosted by high-quality crystalline oxides may be driven by much lower $j_c$ for movement due to low defect-pinning. The present records in amorphous metallic heterostructures ($10^{11}$-$10^{12}$ A/m$^2$) due to high defect-pinning, while those in bulk single-crystalline B20 compounds can be as low as $10^5$-$10^6$ A/m$^2$).
- Perovskite oxides with $t_{2g}$ conduction band or in (111)-orientation have topologically non-trivial and highly spin-textured band structures.
- When Skyrmions driven by large current to move via Eqn. 4, oxides with higher molar heat capacity $C_m$ would produce less temperature rise compared to metals, hence magnetic properties are more stable against thermal fluctuations.

*1.4 Popular imaging techniques for magnetic domains/bubbles/Skyrmions*

The earliest search for dipolar-stabilized bubbles began around 1970s and was usually performed by Faraday Effect[50-52], which is the transmission version of the modern Magnetic Optical Kerr Effect (MOKE), involving a rotation of linear polarization of incident light by a particular magnetic material. This technique showed contrast between the upward and downward magnetization domains but were not sensitive to the intermediate in-plane moment regions, hence information about $Q_h$ of the bubbles was concealed. Likewise, MOKE is popular among researchers working on Néel-Skyrmions[43] hosted by metallic ultrathin films where dark nano-sized dots are seen in a light background. Although the DMI is well-accepted to be interfacial Rashba-type, no topological parameters can be revealed. Nevertheless it is a convenient and non-destructive measurement tool for Skyrmions' motions.

The Lorentz Transmission Electron Microscopy (LTEM) is by far one of the most convincing tools for magnetic bubbles and Skyrmions imaging, with the moment orientations analysed by transport-of-intensity equation (TIE)[53, 54]. LTEM intensity contrast is $I \propto (\nabla \times \hat{m}) \cdot \boldsymbol{k}$, implying that it is sensitive to only a curl of moments around the incident electron beam $\boldsymbol{k}$. Since $\boldsymbol{k}$ is usually parallel to the normal vector of a thin sample plate (thinned by ion milling to <100 nm thickness for sufficient electron transmission), the out-of-plane moments will appear dark but can be inferred by acknowledging that the moments of majority surrounding area is parallel to field while the Skyrmion/bubble core is antiparallel. This way, LTEM can clearly reveal the topological parameters ($Q$, $Q_v$, $Q_h$) and distinguish among mix-helicity (type-1) bubbles, trivial (type-2) bubbles, uniform-helicity Bloch-type Skyrmions, antiskyrmions[55] as well as some intermediate transformations. Examples are given in Table 1. Despite its resolution advantage, LTEM imaging is challenging for Néel-Skyrmions, since Néel-Skyrmions have $\nabla \times \hat{m}$ perpendicular to the sample's normal vector and $\boldsymbol{k}$, thus requiring a sample tilt[56], and their typical ultrathin film hosts do not provide adequate interaction volume for LTEM.

On the other hand, the small-angle neutron scattering (SANS) is the only imaging technique for magnetic textures in the reciprocal-space, and it pioneered the discovery work of a true Bloch-type Skyrmion on year 2009 by Mühlbauer[3]. The SANS patterns can reveal long wavelength modulations of magnetic moments along particular crystal orientation. By analysing the relationship between scattering patterns with the directions of magnetic field ($\boldsymbol{H}$) and incident neutron ($\boldsymbol{k}_n$), the magnetic phases can be deduced. Nevertheless, it is only suitable for dense Skyrmion lattices and bulk samples.

In recent years, more varieties of non-destructive imaging techniques were developed, particularly complementing LTEM's shortcoming on Néel-Skyrmions. The Magnetic Force Microscopy (MFM)[57], Spin-polarized STM (Sp-STM)[16, 28, 29] and Nitrogen-Vacancy Magnetometry (NVM)[58, 59] are the prominent scanning probe techniques. Although Sp-STM offers much higher resolution compared to MFM and has done a great job in imaging atomic scale Néel-Skyrmions such as the one in Fe/Ir(111), its reliance on tunnelling current limits its usage within the class of metallic ultrathin films. NVM is believed to bridge the mentioned compromise between Sp-STM and MFM, but has not been demonstrated. The X-ray Magnetic Circular Dichroism Photoelectron Emission Microscopy (XMCD-PEEM)[60, 61] and Scanning Transmission X-ray Microscopy (STXM)[13, 44] are the high-brilliance X-ray techniques relying on 3$^{rd}$-generation synchrotrons, with magnetic resolution similar to MOKE and MFM. However, we note that the XMCD-PEEM is inapplicable to insulators, since PEEM ejects electron from the sample surface and creates charging effect; while STXM also involves difficulty in fabricating samples onto nm-thick membranes. Lastly, the Spin-polarized Low-energy Electron Microscopy (SPLEEM)[62] is another high resolution Skyrmion imaging technique, capable of resolving the topological details of Néel-Skyrmion[63, 64] akin to the quality achieved by Sp-STM and LTEM. Its usage of low electron energy (few eV) also makes it highly surface-sensitive.

Having established a good understanding on the valuable aspects of Skyrmions, the subsequent sections are dedicated to a thorough review on oxide Skyrmions, which is a broad platform with numerous interesting physical aspects that offer extra freedoms of control.

## 2. Magnetic Bubbles in Centrosymmetric Crystals Stabilized by Dipolar/Demagnetization Field

*2.1 Dipolar or Demagnetization contribution*

At larger scale, a 5$^{th}$ term for dipolar/demagnetization energy $\sum_{i=x,y,z} \frac{\mu_o}{2} N_i M_s^2$ can be added into the energy landscape (Eqn. 2) for materials with negligible DMI due to their centrosymmetric crystal structures, where demagnetization field $H_{d,i} = -N_i M_s$, $N_i$ is the demagnetization factor and $M_s$ is the saturation magnetization. For any thin ferromagnetic film/plate[65], we can consider a cylindrical disc with radius $R$ aligning in z-direction of the plate of thickness $L$, we have $N_z = (L + R - \sqrt{L^2 + R^2})/L$, while $N_{x,y} = (\sqrt{L^2 + R^2} - R)/2L$. Hence it is easy to see that this demagnetization term yields a shape anisotropy that favours in-plane magnetization for thin plates at large limit of $R/L$ where $N_z \to 1$ and $N_{x,y} \to 0$. To produce the vortex-shape magnetic bubbles, large uniaxial magnetocrystalline anisotropy $K_u$ is needed. In fact, these bubbles have surfaced since the 1970s, and they do not always fulfill Eqn. 1 but may result in richer and often uncontrollable varieties of topologies, i.e.: unlike Skyrmions stabilized by DMI, their helicity $Q_h$ is not uniformly single-valued across the whole sample, but with randomly mixed with clockwise (CW) and counterclockwise (CCW)[51, 66] handedness. In oxides, such bubbles were found in manganites, garnets, hexaferrites and orthoferrites in the past few decades. They are well-described by the wall-energy model[51, 52, 67, 68] (note that in Erg. unit: $\mu_o=4\pi \times 10^{-7}$, the energy quantities are divided by $10^{-7}$, $\mu_o$ replaced by $4\pi$):

- Quality factor, $q = \frac{2K_u}{\mu_o M_s^2} > 1$ to produce bubbles (5)
- $\sigma_w = 4\sqrt{J_{ex}K_u}$ is the domain wall energy per unit area
- Characteristic length, $l = \frac{\sigma_w}{\mu_o M_s^2}$
- Optimal plate thickness $h < l$
- Domain wall thickness, $w \sim \frac{l}{q} = \sqrt{\frac{J_{ex}}{K_u}}$
- Bubble size $\approx 8\, l = 32 \frac{\sqrt{J_{ex}K_u}}{\mu_o M_s^2}$
- Note that reducing $h$ can reduce $l$ and also bubbles size.
- Note: If DMI of a particular sign is present, $\sigma_w = 4\sqrt{J_{ex}K_u} - \pi D$ is modified by DMI[17, 69], causing helicity $Q_h$ of one type to be more stable than the other.
- If DMI is present together with demagnetization, then reducing $h$ should cause enhancement in SOC, $K_u$ and q, such that the bubble phase space is enlarged.

## 2.2 Oxides Thin-plate Skyrmions with Mixed Helicities

Several magnetic bubbles stabilized by dipolar/demagnetization were found in the early 1970s. In oxides, these bubbles usually form in the magnetic insulator phases, which unfortunately limited their use in electronics. Here we present a list of dipolar-stabilized oxide magnetic bubble with their characteristics in Table 1, which were mainly studied by LTEM:

| Name | Crystal structure | Orient. | Thick. (µm) | $T_{Curie}$ (K) | $T_{Bubble}$ (K) | $\mu_o H$ (mT) | Bubble Type | Dia. (µm) | Refs. |
|---|---|---|---|---|---|---|---|---|---|
| La$_{0.875}$Sr$_{0.125}$MnO$_3$ | Perovskite | (011)$_o$ | 0.05-0.25 | 180 | 100 | 280-360 | Bloch | 0.1 | [70] |
| La$_{0.825}$Sr$_{0.175}$MnO$_3$ | Perovskite | (001)$_o$ | 0.05-0.15 | 280 | 90-100 | 185-470 | Bloch, type-2 | 0.3 | [71-73] |
| La$_{0.5}$Ba$_{0.5}$MnO$_3$ | Perovskite | Poly-crystals | 0.07 | 300 | 300-360 | 0-150 | Bloch, dumbbell | 0.2 | [74] |
| La$_{1.37}$Sr$_{1.63}$Mn$_2$O$_7$ | Layered perovskite | (001) | 0.1-5 | 100 | 20-46 | 350 | Biskymion | 0.2 | [75, 76] |
| Doped BaFe$_{12}$O$_{19}$ | Hexagonal | (001) | 0.03 | 453 | 300 | 100-200 | Bloch, type-2 | 0.15 | [77, 78] |
| R$_3$Fe$_5$O$_{12}$ | Garnet | various | 11-25 | 570 | RT | 1-37 | No info | 2.5-20 | [79, 80] |
| RFeO$_3$ | Orthoferrites | (001)$_o$ | 28-76 | 645 ($T_N$) | RT | ~0.3-6 | No info | 19-190 | [50] |

**Table 1.** List of oxides hosting dipolar-stabilized bubbles. The abbrevations 'Orient'=crystal plane normal to magnetic field. 'Thick.'=plate thickness, and 'Dia.'=bubble diameter. Here the 'Bloch-type' refers to the proper Skyrmion-like bubbles with Q=1. Type-2 bubbles have no net topological charge. *Data adapted from respective journals with permissions.*

For La$_{1-x}$Sr$_x$MnO$_3$ (LSMO) at low Sr doping, the Jahn-teller distortion of $e_g^1$ configuration from the parent LaMnO$_3$ is still partially active, promoting lifting of Kramer's degeneracy and orbital-ordering (OO), creating the ferromagnetic (FM) insulating OO phase. This OO involving double-exchange among the $d_{x^2-y^2}$ and $d_{3z^2-r^2}$ orbitals possibly contributes large $K_u$ along [001]$_o$[70]. In general, the well-known phase diagram for LSMO consists of:

- 0<x<0.1: canted antiferromagnetic (AF) for $T<T_N$, paramagnetic (PM) insulator for $T>T_N$
- 0.1<x<0.15: FM insulator for $T<T_{MI}$ = 150 K, FM metal for $T_{MI}<T<T_C$, PM insulator for $T>T_C$.
- 0.15<x<0.3: No FM insulator phase, FM metal at $T<T_C$, PM insulator at $T>T_C$. Has a structural phase transition from orthorhombic ($T<T_S$) to rhombohedral ($T>T_S$) where $T_S<T_C$.
- 0.3<x<0.6: FM metal for $T<T_C$, PM metal for $T>T_C$.
- x>0.6: canted AF for $T<T_N$, PM insulator for $T>T_N$

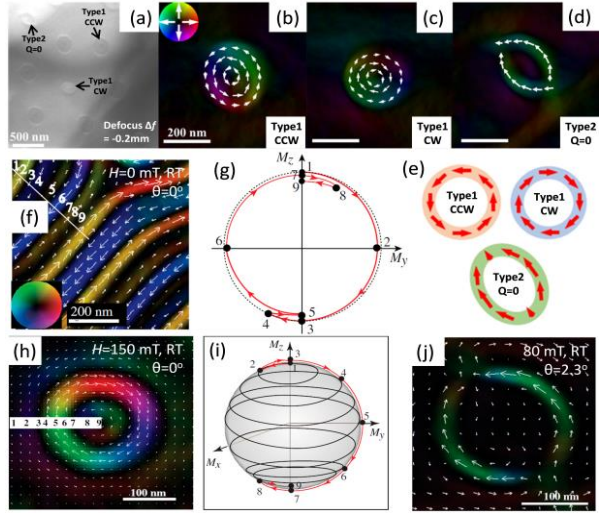

**Figure 2.** (a)-(e) correspond to $La_{0.825}Sr_{0.175}MnO_3$, while (f)-(j) correspond to $BaFe_{10.35}Sc_{1.6}Mg_{0.05}O_{19}$ hexaferrite. (a) Random magnetic bubbles were found in LSMO (x=0.175), which can be analyzed by TIE into three types of bubbles in (b)-(d). (e) Schematics of the in-plane moment alignments in the three types of bubbles. (f) In the hexaferrite, helical stripes are found at zero field, with some $M_z$ oscillations analyzed in (g) according to the number labels. (h) Concentric rings of alternating type-1 bubbles at intermediate out-of-plane fields $H_z$, with some $M_z$ oscillations analyzed in (i) according to the number labels. (j) Tilting the magnetic field by an angle θ away from the out-of-plane causes topological transformation into type-2 bubbles. *Reprinted (adapted) with permission from Ref.* [73, 77]. *Copyright (2016) American Physical Society and (2012) PNAS.*

Generally, the external magnetic field ($H_{ext}$) and imaging electron beam $k$ vector should be aligned along the uniaxial easy axis with the largest $K_u$ to realize stripe domains (at zero-field) and bubble formations (with small mT fields); otherwise, large in-plane collinear domains with narrow domain walls will be found. Increasing $H_{ext}$ too large along the easy axis also causes bubbles to shrink and vanish; this is intuitive since $H_{ext}$ is typically parallel to the circumference moments but antiparallel to the core moments. In LSMO (x=0.175), rising above the structural phase transition temperature ($T_S$ = 190 K), the rhombohedral structure is unable to provide sufficient $K_u$ for bubble formation, and crossing the phase boundary causes abrupt (1$^{st}$ order) disappearance of bubbles. In the orthorhombic phase ($T<T_S$) at (001)$_o$-orientation, besides showing coexistence of mixed $Q_h=\pm\pi/2$ (CW and CCW) due to their degeneracy (Fig. 2a-c), the "type-2 bubbles" with two counter-propagating Bloch walls (which are not Skyrmion, $Q=0$) can also be randomly found (Fig. 2d-e). By tilting the magnetic field slightly away from the easy axis towards in-plane (2°-3°), majority of the proper Bloch-type Skyrmion-like bubbles ($Q=1$) will transform into the type-2 bubbles following some distortion of Bloch lines[71, 72, 77] (Fig. 2j). On the other hand, composite bubbles similar to Biskyrmion also formed but one of the pair has $Q=0$[71].

In $BaFe_{10.35}Sc_{1.6}Mg_{0.05}O_{19}$ hexaferrite[77], although the stripe domains at zero-field generally follow the expected sequence in a Bloch sphere, some repetitions in the form of "oscillation" are spotted around $\pm M_z$ (Fig. 2f-g). Such complexity yielded three concentric CW and CCW rings with $M_\parallel$ in one bubble when a field of 150 mT was applied (Fig. 2h-i). Yet, it is still having $Q=1$ but not a concentric Bloch-Skyrmion of $Q=3$, since in overall the moments only wrap around the Bloch sphere once. Such oscillation in Bloch sphere is also encountered again in LSMO (x=0.175)[71].

In the reported $La_{0.5}Ba_{0.5}MnO_3$ case[74], the polycrystalline random orientation could be the root cause for persistence of Skyrmion-like bubbles up to 360 K which is above $T_{Curie}$. One would note that the moment versus temperature (M-T) curve show significant non-vanishing moment above $T_{Curie}$, similar to magnetic polaron or chirality fluctuation as observed before in $La_{1-x}Ca_xMnO_3$[81, 82] as well as recently in $SrRuO_3$[83] thin films, via as Topological Hall Effect (THE). The author suggested that there should another $T_{Curie}$ (~350 K) accounting for inherent random potential by quenched disorder. In particular, the typical stripe domains expected at zero-field in other afore-mentioned system is absent, but random bubbles appear in a contrast-less ferromagnetic background. Another interesting feature is the coexistence of bubbles with +/-Q signs, in addition to the mixed-sign $Q_h$ (Fig. 3a-d). They were distinguished by their response to sweeping magnetic field, i.e.: one bubble showed shrinkage of circumference by sweeping field from 0 to -100 mT (Fig. 3a-b), while the other bubble showed core expansion by sweeping field from 0 to -150 mT (Fig. 3c-d), where both field are downward. This could only imply the former has downward circumference moments and upward core moments, while the latter is exactly opposite. We speculate that such spontaneous $Q$ sign-reversal at a fixed field could be very common due to thermal fluctuation and low coercive field ($H_c$) near $T_{Curie}$, but should be absent at $T \ll T_{Curie}$ and large $H_c$.

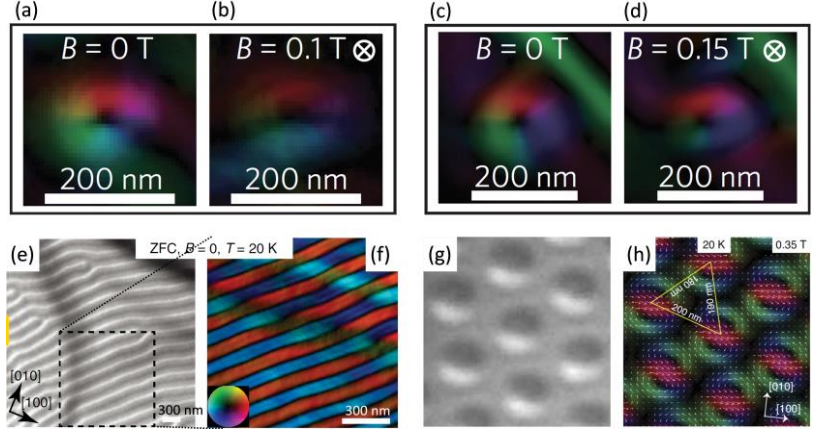

**Figure 3.** (a)-(d) correspond to $La_{0.5}Ba_{0.5}MnO_3$ while (e)-(h) correspond to $La_{1.37}Sr_{1.63}Mn_2O_7$. With the application of a downward field, a core expansion in (a)-(b) but a circumferential shrinkage in (c)-(d) can be seen. (e)-(f) Helical stripe domains at zero-field transformed into Biskyrmions in (g)-(h) after applying a small $H\|[001]$. (f) and (h) are TIE-analysis results from (e) and (g). A Biskyrmion in (g)-(h) can be understood as a composite formed by a pair of type1-CCW and -CW bubbles sharing a common Bloch wall, due to lack of DMI. *Reprinted (adapted) with permission from Ref.* [74, 76]. *Copyright (2013) and (2014) Nature Publishing Group.*

In $La_{1.37}Sr_{1.63}Mn_2O_7$ with tetragonal structure and easy axis [001] studied by X.Z. Yu[76], biskyrmions were unambiguously spotted (Fig. 3g-h). In Biskyrmions, two Q=1 and but opposite-sign $Q_h=\pm\pi/2$ bubbles are merged together into a Q=2 pair, sharing a common Bloch wall. In fact this is only possible when DMI is absent, and dipolar field ensures that $Q_h=\pm\pi/2$ are degenerate. As a counterexample, two bubbles with $Q_h$ of the same sign cannot share an oppositely oriented Bloch wall. Besides, type-2 (Q=0) bubbles also emerge at zero-field if it is first subjected to a field-cooling (FC) procedure with a field of 150 mT. An attraction force between the Bloch walls of nearby CW and CCW helicity will align the Biskyrmions in forming polar single domain lattice everywhere. One interesting feature occurs when the field is swept back to zero from 0.35 mT without reaching the annihilation field of 400 mT, the Biskyrmions will co-exist with stripes. Another amazing work in that paper demonstrated Biskyrmion flow at an encouragingly low current density of 7 x $10^7$ A/m$^2$, where the L-TEM images for Biskyrmions were blurred, but the stripes image still remains sharp and not moving. Hence, this again verified the intrinsic property of Skyrmion having less pinning and lower $J_C$ compared the helical stripes, consistent to Eqn. 4 and the finding of Ref. [40].

## 3. Multiferroic Skyrmions in $Cu_2OSeO_3$

### 3.1 Basic Properties

The multiferroic $Cu_2OSeO_3$ single crystals are usually grown by chemical vapour transport with HCl carrier gas[84]. $Cu^{2+}$ with electronic configuration 3d $t_{2g}^6 e_g^3$ (S=1/2) is the only species that contributes magnetic moment with 0.5 $\mu_B/Cu^{2+}$ at (high-field) saturation. In a $Cu_2OSeO_3$ unit cell, $Cu^{2+}$ ions are surrounded by square-pyramidal or trigonal bi-pyramidal oxygen cages in ratio of 3:1, yielding a three-up-one-down ferrimagnetic system (Fig. 5a)[12]. The crystal structure is non-centrosymmetric and contributes a strong Dresselhaus-type DMI.

Using SANS on bulk $Cu_2OSeO_3$ crystal, T. Adams et al. realized that its magnetic ground state is not simply ferrimagnetic, but helimagnetic[85]. In SANS (Fig. 4a), the multi-$q$ helical phase produces a fourfold or twofold symmetric scattering pattern, with modulation vector perpendicular to the magnetic field ($q\perp H$). The single-$q$ conical phase produces a twofold symmetric pattern, but its distinct feature is $q\|H$. Whereas a triple-$q$ Skyrmion lattice would produce a unique six-fold symmetric pattern when the field and incident neutron are parallel but perpendicular to the modulation vectors ($H\|k_n\perp q$). Generally, by subjecting magnetic field onto any crystallographic direction of $Cu_2OSeO_3$, a common phase diagram would be obtained (Fig. 4b): i.e. at temperature far below $T_{Curie}$, helical stripes (with modulation wavelength, $\lambda_h$ of 50-60nm) dominate the zero- to low-field regime; the conical phase exists at the intermediate-field regime; and the collinear ferrimagnetic phase exists at high-field. At temperature slightly below $T_{Curie}$, the Skyrmion A-phase can be found at the intermediate-field regime, housed within the conical phase. Adams et al. further discussed that although the temperature-dependence of the helical phase in $Cu_2OSeO_3$ shows a 2$^{nd}$-order phase transition, its specific heat capacity shows a 1$^{st}$-order phase transition with a latent heat peak (Fig. 4c). This indicates the onset of helical spin fluctuation at slightly above $T_{Curie}$, and is likely the reason of enhanced magneto-dielectric response (MDR), thus magnetoelectric (ME) coupling.

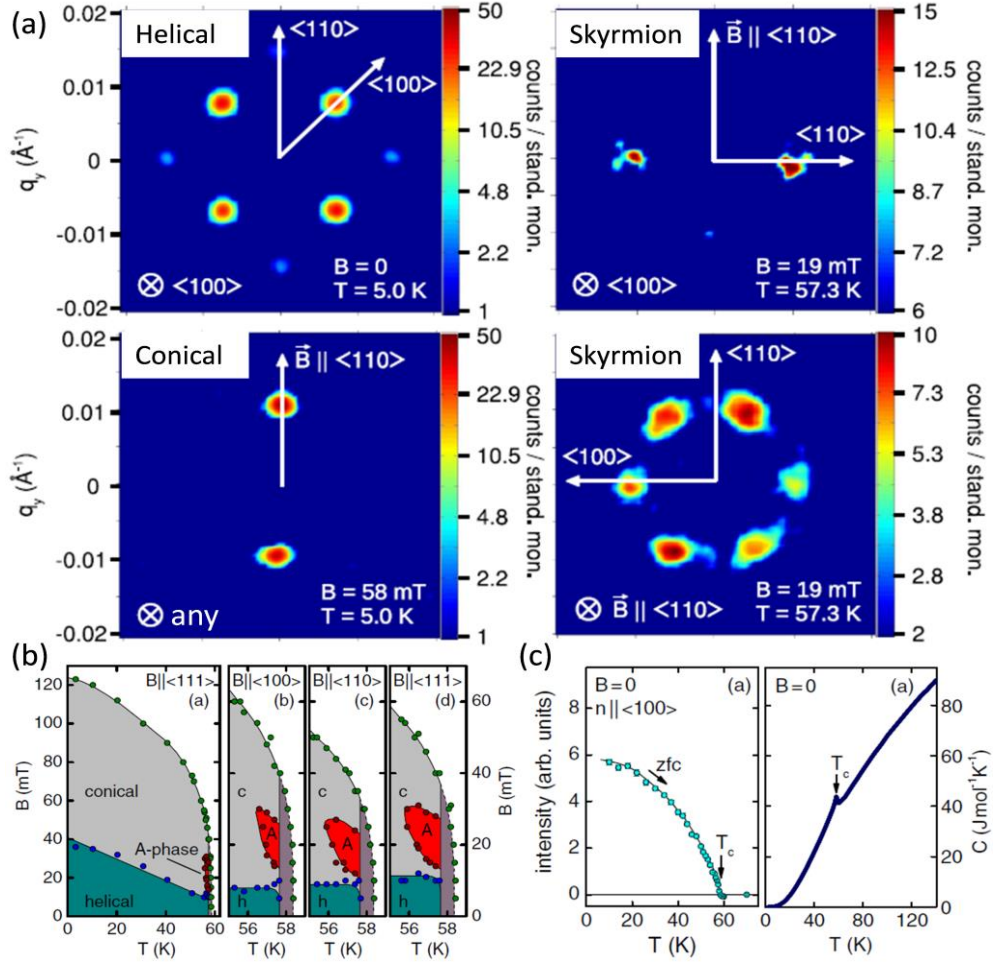

**Figure 4.** (a) Characteristics of SANS pattern for helical, conical and Skyrmion phases. (b) Common phase diagrams with magnetic field applied onto various crystallographic directions, where h, A, c denotes helical, Skymion and conical phases. (c) Temperature dependence of SANS intensity (left) and specific heat capacity (right) in the helical phase. *Reprinted (adapted) with permission from Ref.* [85]. *Copyright (2012) American Physical Society.*

The seminal work of LTEM imaging on $Cu_2OSeO_3$ was performed by Seki et al. with $H$∥[110] and [111] (Fig. 5b-e). Since DMI is active, these Skyrmions have uniform $Q_h$, unlike the dipolar-stabilized ones discussed in section 2. $Cu_2OSeO_3$ in thin film form was also found to exhibit Skyrmion phase space enlargement behaviour (Fig. 5f) consistent to other B20 compounds[86]. The $P2_13$ space-group (threefold rotation around [111]) should be nonpolar, but the presence of $M_{[111]}$ removes this symmetry and creates a net $P_{[111]}$. Hence, the ME coupling of $Cu_2OSeO_3$ can be characterized by AC magnetometry, which is able to detect frequency-dependent resonances. As seen in Fig. 5g-h, for the bulk $Cu_2OSeO_3$ at any temperature below $T_{Curie}$ and zero-field, the electric polarizations in the multi-$q$ stripe domains are randomly-aligned and cancel out ($P$=0). Increasing $H$∥[111] causes $P_{[111]}(H)$ to increase in negative direction and reverse sign to positive, which can be fitted by a parabolic $P \propto M^2$ trend into the conical phase. Finally, $P_{[111]}$ saturates when $M_{[111]}$ saturates into collinear ferrimagnetic. In the Skyrmion A-phase within phase space of 56-58 K and ~20-40mT, $P_{[111]}(H)$ shows a hump anomaly. Correspondingly, the AC susceptibility $\chi' = \frac{\partial M}{\partial H}$ also shows a dip anomaly in both the helical and Skyrmion phase spaces (Fig. 5h).

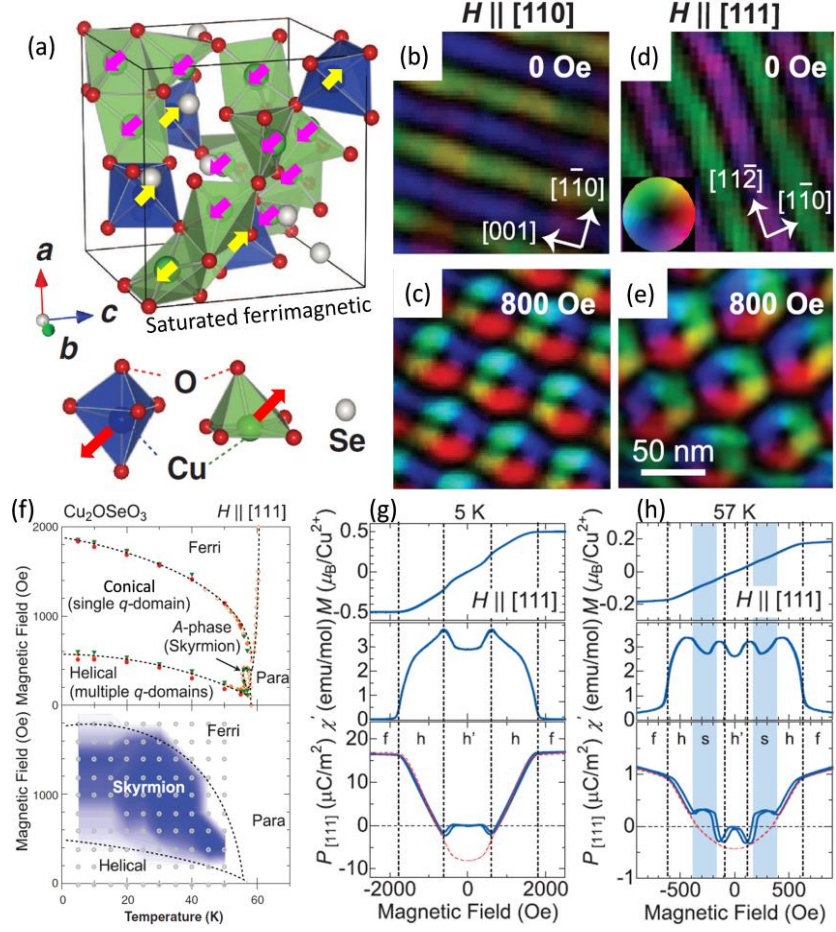

**Figure 5.** (a) Crystal structures and $Cu^{2+}$ (S=1/2) magnetic moment directions of $Cu_2OSeO_3$ at the saturated ferrimagnetic state at large $H$∥[111]. The atoms and oxygen cages are labelled at the bottom panel. Zero-field multi-*q* helical stripe domains and low-field Skyrmions phases at $H$∥[110] (b)-(c) and $H$∥[111] (d)-(e) respectively. (f) Magnetic phase diagram for bulk (top) and thin film (bottom) respectively. With [111]-oriented magnetic field for the bulk crystal, the magnetization (top), AC susceptibility (middle) and electric polarization (bottom) were measured at (g) 5 K (Skyrmion absent) and (h) 57 K (Skyrmion present) temperatures respectively. *Reprinted (adapted) with permission from Ref. [12]. Copyright (2012) Science AAAS.*

*3.2 Magnetoelectric Coupling, Vibrational Modes, and Metastable Skyrmion Phase*

The mentioned parabolic ***P*-*M*** relationship agrees with the d-p hybridization model[87, 88] – at a cation-anion pair i-j where the moment residing at the cation site i, the polarization is related to the moment by $\boldsymbol{P}_{ij} \propto \left(\boldsymbol{e}_{ij} \cdot \boldsymbol{M}_i\right)^2 \boldsymbol{e}_{ij}$ where $\boldsymbol{e}_{ij}$ is the bonding direction. Using angle-dependent ***M*** and ***P*** measurements, Seki et al. verified that ***H***∥[111] induces $\boldsymbol{P}_{[111]}$, ***H***∥[110] induces $\boldsymbol{P}_{[001]}$, while ***H***∥[001] causes ***P***=0 in all directions (Fig. 6a), which is consistent to $(\boldsymbol{P}_i, \boldsymbol{P}_j, \boldsymbol{P}_k) \propto (\boldsymbol{m}_{010}\boldsymbol{m}_{001}, \boldsymbol{m}_{001}\boldsymbol{m}_{100}, \boldsymbol{m}_{100}\boldsymbol{m}_{010})$[89]. Such ME coupling rule by symmetry as well as the parabolic relationship are applicable for all the helical, conical, Skyrmion and collinear ferrimagnetic phases, except small differences in coefficient exist in the parabolic relationship between the phases. Seki et al. also calculated detailed mappings of polarization and charge density of a Bloch-type Skyrmion in different crystallographic orientations (Fig. 6b-h), which is beneficial for designing electric-field tunable Skyrmions[87].

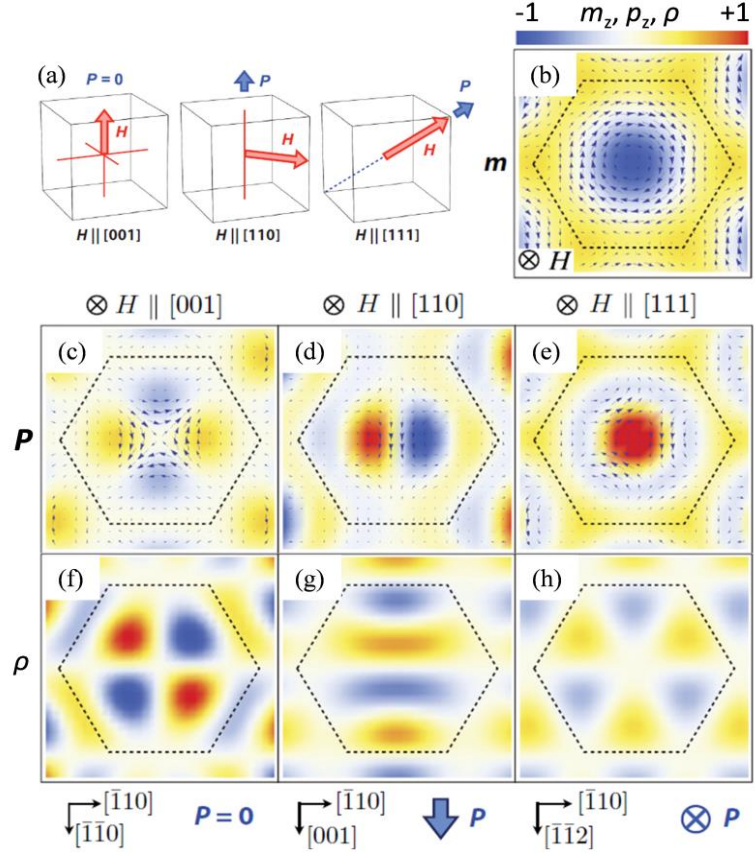

**Figure 6.** (a) Summary of ME coupling directions in a $Cu_2OSeO_3$ unit cell. (b-g): Mappings of out-of-plane components of moment *m* (b), electric polarization *P* (c-e) and charge density $\rho$ (f-h) corresponding to three magnetic field directions with respect to crystallographic orientations when the Skyrmion A-phase is nucleated. Dotted lines border a magnetic unit cell of a single Bloch-type Skyrmion. *Reprinted (adapted) with permission from Ref.* [87]. *Copyright (2012) Nature Publishing Group.*

Next, the Skyrmion vibrational modes (Fig. 7a-c) in $Cu_2OSeO_3$ can be probed by using ferromagnetic resonance (FMR). The geometry used by Onose et al. was a thin plate with surface normal at [$\bar{1}$10] and an oscillating AC magnetic field ($H_{AC}$) of the microwave is aligned with the in-plane [110] direction[90]. A static DC field was then applied either onto [$\bar{1}$10] or [110], hence forming the $H_{DC} \perp H_{AC}$ or $H_{DC} \parallel H_{AC}$ configurations. At out-of-plane $H_{DC} \parallel [\bar{1}10]$, 40 mT is needed for saturation into collinear, but only 20 mT is needed for saturation with in-plane $H_{DC} \parallel [110]$, because a thin plate has shape anisotropy that favours in-plane easy axis and stronger demagnetization field at out-of-plane. At 57.5 K, Skyrmion-lattices are created at $H_{DC} \parallel [\bar{1}10]$ within 14-32 mT and $H_{DC} \parallel [110]$ with 5-15 mT, as judged from $\chi'(H)$ dip anomalies. Correspondingly, the $H_{DC} \perp H_{AC}$ configuration showed resonance peaks at 1.8 GHz for the multi-*q* helical stripe domains (near $H_{DC}$=0) and conical phase ($H_{DC}$=34-40 mT), 1 GHz in the Skyrmion phase ($H_{DC}$=14-32 mT), and 2.5 GHz in the collinear phase ($H_{DC}$=100 mT) (Fig. 7e). Whereas for the $H_{DC} \parallel H_{AC}$ configuration, the only resonance is at 1.5 GHz in the Skyrmion phase ($H_{DC}$=5-15 mT) (Fig. 7f). The 1GHz resonance at $H_{DC} \perp H_{AC}$ and 1.5GHz at $H_{AC} \parallel H_{DC}$ are the counter-clockwise (CCW) sloshing rotation and breathing modes of Skyrmion excitations respectively.

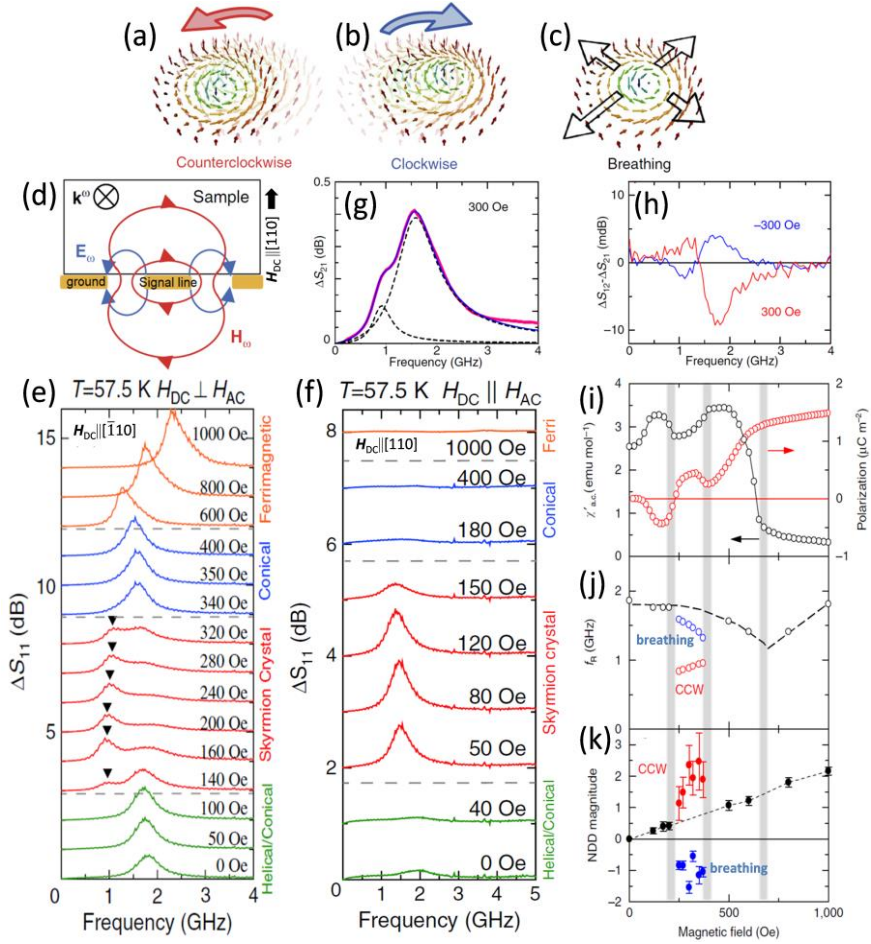

**Figure 7.** (a) CCW-, (b) CW- sloshing rotations and (c) breathing modes of Bloch-type Skyrmions under an AC magnetic field perturbation in additional to the DC magnetic field. The CCW and CW rotations require $H_{AC} \perp H_{DC}$ but the breathing mode requires $H_{DC} \parallel H_{AC}$. (d) The coplanar waveguide (CPW) setup employed in (g)-(k). Microwave absorption spectra performed by Onose et al. revealing the CCW rotation by (e) $H_{AC} \perp H_{DC}$ and breathing modes by (f) $H_{DC} \parallel H_{AC}$ configurations respectively. (e) Both the CCW rotation and breathing modes were co-detected using the CPW setup, but opposite-sign NDD spectra. (f). Magnetic-field dependence of the AC $\chi'$ and polarization (i), FMR frequencies (j) and NDD (k) are summarized. *Reprinted (adapted) with permission from Ref.* [90, 91]. *Copyright (2013) American Physical Society and Nature Publishing Group.*

While Okamura et al. employed a coplanar waveguide (CPW) (Fig. 7d) that is able to access both the $k_{[1\bar{1}0]}(\omega) \parallel (P_{[001]} \times M_{[110]})$ and $k_{[001]}(\omega) \perp (P_{[001]} \times M_{[110]})$ setups, where $k$ is the microwave propagation vector, while the $H_{DC} \parallel [110]$ is fixed[91]. Within each setup, both $H_{AC} \parallel [110]$ and $H_{AC} \parallel [001]$ exist, which is an advantage of the compact CPW. Particularly, Okamura investigated the Nonreciprocal Directional Dichroism (NDD) = $\Delta S_{12} - \Delta S_{21}$ where $\Delta S_{ij}$ and $\Delta S_{ji}$ are the absorption coefficients at opposite $k$-vectors. This is based on the concept that ME coupling causes the absorption intensity of linearly polarized light to depend on constructive or destructive interference. Firstly, in the collinear phase at large $H_{DC}$, NDD peak for the $k_{[1\bar{1}0]}(\omega) \parallel (P_{[001]} \times M_{[110]})$ setup at 1.8GHz shows sign-reversal from positive to negative upon reversing the direction of $H_{DC}$ from positive to negative. This is because upon reversing the signs of $H_{DC}$ and $M$, $(P \times M)$ should have sign-reversal, which can be understood from the fact that $P_{[111]}(H_{DC}) \propto M^2$ discussed earlier is a parabolic even function and does not reverse sign. However, the NDD for the $k_{[001]}(\omega) \perp (P_{[001]} \times M_{[110]})$ is always zero and unresponsive to $H_{DC}$ sign-reversal, which is expected from the selection rule of microwave absorption. Secondly, in the Skyrmion phase, absorption for the $k_{[1\bar{1}0]}(\omega) \parallel (P_{[001]} \times M_{[110]})$ setup showed two close broad peaks at 1 GHz (NDD>0) and 1.5 GHz (NDD<0). These two peaks correspond to the CCW sloshing rotation and breathing modes of Skyrmions respectively (Fig. 7g), and their co-detection is because both $H_{AC} \parallel [110]$ and $H_{AC} \parallel [001]$ components exist in CPW. Probing such NDD response (Fig. 7h, k) is a direct evidence of ME coupling.

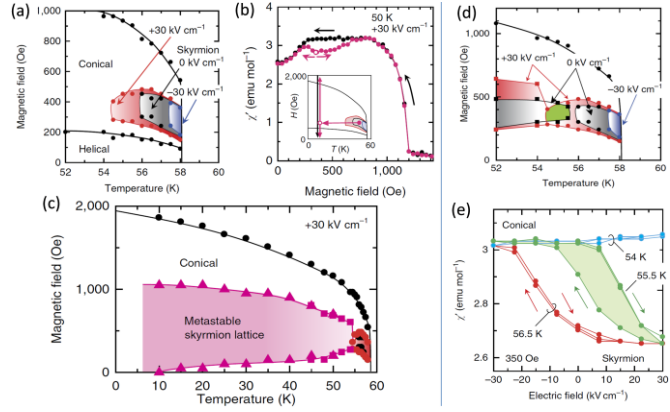

**Figure 8.** (a) Magnetic phase diagram showing the enlargement or shrinkage of normal Skyrmion A-phase by positive and negative electric fields respectively. (b) The ME cooling scheme (35 mT and +30 kV cm$^{-1}$) following the magenta line is able to recover the χ' dip anomaly at 50 K, implying the enlargement of normal Skyrmion phase space; but it is not possible by cooling with large *H*-field or absence of *E*-field (black-line). The two cooling schemes are depicted in the inset of (b). (c) Phase diagram of metastable Skyrmions (magenta) induced by ME-cooling. (d) After ME cooling, removal or polarity switching of E-field causes partial breaking of phase space between normal and metastable Skyrmions. (e) The green-shaded region in (d) can be used for non-volatile switching between Skyrmion and conical phases. *Reprinted (adapted) with permission from Ref.* [92]. *Copyright (2016) Nature Publishing Group.*

After confirming ME coupling, Okamura et al. attempted ME cooling with both magnetic and electric field, induced a metastable Skyrmion phase in bulk $Cu_2OSeO_3$[92]. The setup was $H_{DC}||E_{DC}||P_{[111]}$ to access the strongest ME coupling, which is expected to tune the uniaxial anisotropy by $\Delta K_u \propto -P_{[111]} \cdot E_{[111]} \propto -m_{[111]}^2 E_{[111]}$, which can be a form of VCMA. Hence a positive electric field will enhance $K_u$, thus enlarge the Skyrmion phase space (Fig. 8a). The suitable ME cooling scheme (Fig. 8b) is by using a small *H*-field (35 mT) within the bulk A-phase Skyrmions phase and E-field of +30 kV/cm, and the metastable Skyrmion phase was shown to persist down to low temperature of ~10 K (Fig. 8c). This is evidenced from the presence (and absence) of the 1.6 GHz microwave absorption peak corresponding to CCW Skyrmion sloshing rotation mode when ME cooling is applied (and removed), while the another absorption peak at 2.8 GHz corresponds to the conical phase. After creating the metastable phase, the phase continuity between the metastable and normal Skyrmions can be broken, starting from ~54.5 K, by switching off or reversing the polarity of the electric field (Fig. 8d). The phase-space overlapped between the metastable Skyrmion at zero electric field and normal Skyrmion at +30 kV/cm at ~55 K is thus identified as the region where the non-volatile switching between Skyrmion and conical phases can be demonstrated (Fig. 8e).

As a closing remark for this section, $Cu_2OSeO_3$ remains the only host for multiferroic Skyrmion thus far, and its complex magnetic phase diagram has continued to produce new Physics discovery. However, its low $T_{Curie}$ and the difficulty in making device integratable thin films would limit its practical application. On the other hand, the multiferroic $BiFeO_3$ with cycloidal antiferromagnetic modulation wavelength of ~64 nm, high Néel temperature ~643 K and high ferroelectric Curie temperature ~1083 K would likely be the next candidate in search of multiferroic Skyrmion[93-95].

## 4. Unconventional Skyrmions in Bulk Crystals

This section introduces two peculiar examples that challenge the usual belief that all bulk non-centrosymmetric crystals should host Bloch-type Skyrmions. Firstly, in general, the crystal structures with $C_{nv}$ symmetry are polar and would satisfy the criterion on hosting Néel-type Skyrmions in bulk crystals, such as $GaV_4S_8$ ($C_{3v}$)[96] with $T_{Curie}$~13 K and Skyrmion phase at 11.6 K, as well as $VOSe_2O_5$ ($C_{4v}$)[97] with $T_{Curie}$~8 K and Skyrmion phase at 7.5 K. Their magnetic phase evolutions in *H-T* phase space is similar to the phase diagrams of commonly-known Bloch-type Skyrmions, but with "cycloidal" replacing "helical", while the conical phase is absent. In SANS for $VOSe_2O_5$, the incident neutron direction is $k_n||$c-axis. The polar c-axis of the material has DMI vector $D\perp$c causes orientation confinement of the magnetic moment modulation vector $q$ onto the in-plane. At ZFC and zero-field, SANS pattern is fourfold symmetric but slightly blurred, implying multi-*q* domains. After field training $H_t||$a=12 mT and SANS measurement at 0 mT, twofold symmetric sharp SANS pattern appeared, which is understood as ordered cycloidal stripes. Both the cycloidal stripes at zero-field and Néel-Skyrmion phase at $H||$c = 1.5 mT show valleys/depressions in AC magnetometry χ'(*H*) similar to $Cu_2OSeO_3$, as well as the imaginary part χ"(*H*). A slight difference in SANS patterns between the Néel-Skyrmion phase in $GaV_4S_8$ and $VOSe_2O_5$ can be noted: in $GaV_4S_8$, clear fourfold and sixfold patterns are separated at $H||$[001] and $H||$<111> respectively (Fig. 9a); but in $VOSe_2O_5$, a superposition of one set of fourfold-symmetric, and two sets of sixfold-symmetric (30°-offsetted) patterns can all be found at $H\|[001]$ (Fig. 9c-d). Again, difficult synthesis via chemical vapour transport and low $T_{Curie}$ limits their practical application. Due to the insensitivity of L-TEM towards Néel-type Skyrmion's topology as mentioned at section 1.4, real-space imaging required low-temperature magnetic force microscopy (MFM)[96] (Fig. 9b). However, this material class offers a good understanding how Rashba-type DMI works in polar materials, as derived earlier by Bogdanov and Yablonskii[9].

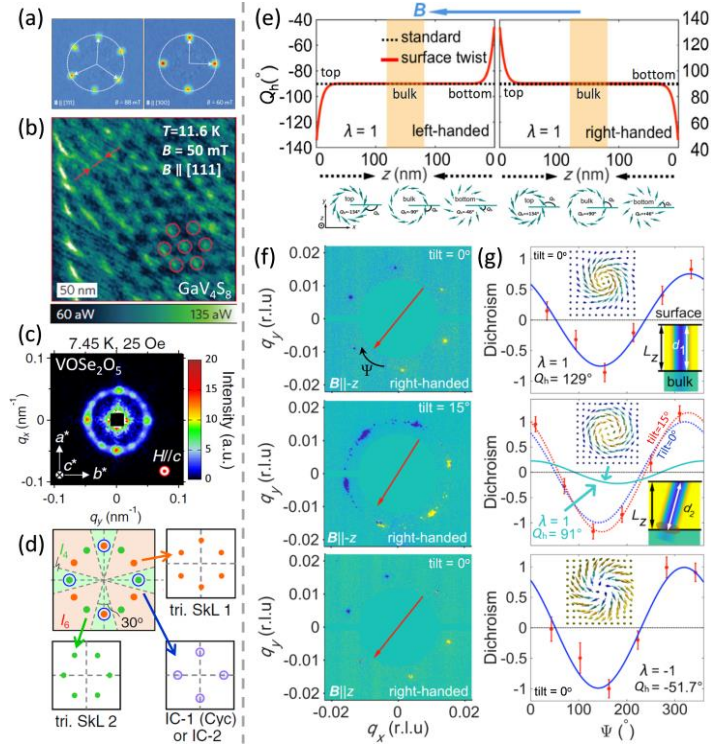

**Figure 9.** SANS pattern (a) and MFM image (b) of Skyrmion phase in $GaV_4S_8$, where the colour scale of (b) represents magnetic interaction energy. (c) SANS pattern of Skyrmion phase in $VOSe_2O_5$, interpreted as a superposition of three magnetic textures in (d), where 'tri', 'SkL', 'IC-1' and 'IC-2' denote 'trigonal', 'Skyrmion-lattice', 'cycloidal' and 'square Skyrmion lattice' respectively. (e) Calculated depth profile of helicity in left- and right-handed $Cu_2OSeO_3$. CD-REXS data (f) and their fitting analyses (g) for the right-handed $Cu_2OSeO_3$ Skyrmion, where the field-inversion can reveal the surface twist at the top and bottom surfaces (top and bottom panels). Middle panel: the bulk state is represented by the cyan curve after subtraction. Inset of (g) indicates the length of Skyrmion cylinders ($d_2>d_1$) being probed. *Reprinted (adapted) with permission from Ref. [96-98]. Copyright (2015) Nature Publishing Group and (2017-2018) American Physical Society.*

Secondly, the broken inversion symmetry at the surface of a non-centrosymmetric crystal structure would contribute an additional $C_{nv}$ symmetry and create a chiral surface twist/reconstruction, resulting in gradual transformation or intermediate state between Néel- ($Q_h=0,\pi$) and Bloch-type ($Q_h=\pm\pi/2$) Skyrmion along its depth profile. This is true if the sample is made thinner than the critical thickness of $8.17\lambda_h$, and is also another mechanism of Skyrmion phase-space enlargement in thin films of B20 compounds. This was first discussed by Rybakov[99] and was observed directly by Zhang et al. in $Cu_2OSeO_3$ using circular dichroism in resonant elastic X-ray scattering (CD-REXS)[98]. The resulted CD-REXS pattern on a Skyrmion lattice is also six-fold symmetric, reminiscent to that of SANS; but the CD bisects the pattern into +/- signs, enclosing a zero-CD vector (extinction direction) that may rotate following the Skyrmion's $Q_h$. Zhang et al. first used theoretical calculations to predict that $Q_h$ approaches 134° and 46° respectively at the top and bottom surfaces of the $Cu_2OSeO_3$ thin plate for a particular out-of-plane magnetic field direction ($B_Z$) (Fig. 9e). This agrees very well with the observed CD-REXS result, where upon the reversal of $B_Z$, the surface $Q_h$ changed from 129° to -51.7° (Fig. 9f-g, top and bottom panels). The $B_Z$ sign-reversal is equivalent to flipping the crystal plate upside down, since in reality the CD-REXS signal is averaged over their limited X-ray penetration depth ($L_Z\sim98$ nm) and is unable to detect $Q_h$ from purely the bottom surface. Zhang et al. also managed to extract and verify the expected bulk $Q_h=91°$ (Fig. 9g, middle panel), which is located far away from both surfaces. This is achieved by first recognizing that $L_Z \gg L_P$ where $L_P\sim7.1$ nm is the decay length of the chiral surface twist, and tilting the magnetic field slightly away (15°) from the surface normal while keeping the incident X-ray at normal direction. This way, the Skyrmion cylinder is tilted, and a higher fraction of the bulk signal will be picked up by the same $L_Z$; and finally a subtraction between the tilted and normal incidence signals yielded the pure bulk $Q_h$.

## 5. Hall Effects in Magnetic Materials (Interlude)

Topological Hall Effect (THE) arises from spin-polarized conduction electrons deflected off non-coplanar magnetic moment textures. A simple picture of this process is that during the hopping process across non-coplanar magnetic textures, conduction electrons have to constantly adjust and align their spin reference frame to the local magnetic moment due to strong Hund's coupling between the free spin and moments ($-J_H \hat{\sigma}_i \cdot \hat{m}_i$), hence gaining a net real-space geometric phase. In the low hole-doped manganites, at the phase transition region from ferromagnetic metal to paramagnetic insulator, hump-shape Hall Effect emerged. Ye[100], Chun[82] and Calderon[101] postulated the existence of Skyrmion strings or chirality fluctuation, using the language of second quantization for tight-binding model:

$$\widetilde{H} = -t \sum_{i \neq j} \cos\left(\frac{\theta_{ij}}{2}\right) e^{i\left[\frac{\phi_{ij}}{2} + \frac{ea}{\hbar c}A_i\right]} c_i^\dagger c_j + \lambda_{soc}(\mathbf{k} \times \nabla V) \cdot \hat{\sigma} + g\mu_B \mathbf{H} \cdot \sum_i \mathbf{m}_i \quad (6)$$

Where $c_i^\dagger$ and $c_i$ are creation and annihilation operators, $\cos(\theta_{ij}) = \mathbf{m}_i \cdot \mathbf{m}_j$, $\mathbf{B} = \nabla \times \mathbf{A}$ is the external magnetic field, and the phase factor $\phi = 2\tan^{-1}\left(\frac{\hat{m}_i \cdot (\hat{m}_j \times \hat{m}_k)}{1 + \hat{m}_i \cdot \hat{m}_j + \hat{m}_j \cdot \hat{m}_k + \hat{m}_k \cdot \hat{m}_i}\right)$ contains the Pontryagin charge. The two phase terms $\frac{\phi(i,j)}{2} + \frac{ea}{\hbar c}A_i$ correspond to THE and ordinary Hall Effect (OHE), while the 2nd and 3rd terms account for SOC and Zeeman Effect respectively. We can understand that an effective magnetic field $\mathbf{b}_{i,\text{eff}} = \nabla \times \nabla \phi = \frac{\epsilon_{ijk}}{4}\hat{m} \cdot (\partial_j \hat{m} \times \partial_k \hat{m})$ emanates from the magnetic textures. This produces the total Hall resistivity $\rho_{xy} = (B + b_{\text{eff}})/ne$ showing the equivalent form of OHE and THE. On the other hand, Bruno[102] and K. Everschor-Sitte[103] used a simpler Schrodinger Hamiltonian for electrons on a topological magnetic textures and unitary transformation for the strong Hund's coupling, and arrived at the same insights for THE. A spin-polarization $P_s = \frac{\text{DOS}_\uparrow(\epsilon_F) - \text{DOS}_\downarrow(\epsilon_F)}{\text{DOS}_\uparrow(\epsilon_F) + \text{DOS}_\downarrow(\epsilon_F)}$ correction factor was also added into $\rho_{\text{THE}}$, where $\text{DOS}_{\uparrow\downarrow}(\epsilon_F)$ is the spin-dependent density-of-states at Fermi level.

For the Anomalous Hall Effect (AHE) that requires out-of-plane magnetization, three distinct mechanisms have been discussed[104-108]. The intrinsic k-space Berry curvature effect can be understood from the semiclassical approach $v_{n,y} = -\frac{i}{\hbar}[eEx, y] = -\frac{e}{\hbar}[\mathbf{E}_x \times \mathbf{\Omega}_{n,z}(\mathbf{k})]$, where $\mathbf{\Omega}_n(\mathbf{k}) = \nabla_\mathbf{k} \times \langle u_n(\mathbf{k})|\nabla_\mathbf{k}|u_n(\mathbf{k})\rangle$ is the Berry curvature, $u_n(\mathbf{k})$ is the Bloch wave function, $x$- and $y$-directions are parallel and perpendicular to the applied electric field $E$ respectively. For side-jump mechanism, Nozières-Lewiner[109] described SOC-assisted electron scattering off an impurity potential $V$ yields a transverse velocity $v_y = \frac{\lambda_{\text{soc}}}{\hbar}(\nabla V \times \hat{\sigma})$. Whereas the skew scattering across an impurity is usually expressed via Fermi golden rule – the transition probability between $|\mathbf{k}, s\rangle$ and $|\mathbf{k}', s'\rangle$ states is asymmetric for left- and right-deflections when SOC is taken account: $W_{|\mathbf{k},s\rangle \to |\mathbf{k}',s'\rangle} = \frac{2\pi}{\hbar}|\langle \mathbf{k}, s|V_{im}|\mathbf{k}', s'\rangle|^2 \delta(\epsilon_{\mathbf{k},s} - \epsilon_{\mathbf{k}',s'})$ where $\langle \mathbf{k}, s|V|\mathbf{k}', s'\rangle = V_{\mathbf{k},\mathbf{k}'}\left(\delta_{s,s'} + \frac{i\hbar^2}{4m^2c^2}(\langle s|\hat{\sigma}|s'\rangle \times \mathbf{k}') \cdot \mathbf{k}\right)$. Typical experiments scale AHE conductivity with mean free time $\tau$. For both intrinsic and side-jump, no dependence on $\tau$ is expected, hence $\rho_{xy} \approx \frac{\sigma_{xy}}{\sigma_{xx}^2} \propto \rho_{xx}^2$; yet for skew scattering, $W_{\mathbf{k} \to \mathbf{k}'} \propto \tau^{-1}$ leads to $\rho_{xy} \propto \rho_{xx}$.

For all three mechanisms of AHE, SOC is directly involved in electron deflection, but is not necessarily involved to produce THE when magnetic textures are present[110]. Ideally, for a magnetic material with out-of-plane uniaxial anisotropy and hosting hexagonal-closed packed magnetic Skyrmion-lattice, the total Hall resistivity is:

$$\rho_{xy}(B) = R_o B + (\alpha \rho_{xx} + \beta \rho_{xx}^2)\lambda_{soc}M + R_o P_s \frac{Qh}{e\pi r_{sk}^2} \quad (7)$$

Where the 1st, 2nd and 3rd terms are OHE, AHE, and THE respectively, $R_o = 1/ne$, and α, β are fitting parameters. The above picture for THE is accurate when $J_H$ is strong and electron hopping is adiabatic (slow), providing adequate time for the fast $\hat{\sigma}$-$\hat{m}$ interaction. Nakazawa recently improvised the THE theory by accounting for non-adiabatic and/or nonlocal $\hat{\sigma}$-$\hat{m}$ interactions[111]. This way, THE behaviour at strong electron-electron correlation and high effective mass (especially true in oxides) can be more accurately modelled. It is customary to understand the field-dependence of $\rho_{\text{OHE}}$, $\rho_{\text{AHE}}$ and $\rho_{\text{THE}}$. For $\rho_{\text{AHE}}(H)$, hysteretic square loops can be fitted with Langevin function: $\mathcal{L}(H) \propto \left\{\coth\left[\frac{g\mu_0\mu_B J}{k_B T}(H \pm H_c)\right] - \frac{k_B T}{g\mu_0\mu_B J(H \pm H_c)}\right\}$ which usually follows the shape of magnetization-field loops. Whereas $\rho_{xy}^{\text{THE}}(H)$ should follow the phase evolution of Skyrmions or bubbles – typically stripe domains exist and no Skyrmion at zero-field; Skyrmions would reach the smallest size and densest lattice packing at intermediate field $H_{peak}$, and finally vanishes into collinear ferromagnet at large field. Hence $\rho_{THE}(H) \sim \frac{L(H)}{1+(H/D_{eff})^2}$ showing hump/dome feature is a good phenomenological fit[83, 112].

THE also appears in pyrochlore frustrated antiferromagnets[113-115]. Theoretical advancements in this field have enabled several insightful Monte Carlo simulations[112, 116, 117]. In the next section, we see that the hump-shape Hall features in SrRuO$_3$ ultrathin films have gained heated debates over its representation of true THE signal or partial cancellation of two oppositely-signed AHE loops. Nevertheless, if without proper imaging, the Hall Effects do not distinguish between the Skyrmions'/bubbles' types, if any of them exists at all.

## 6. Topological Hall Effect and Skyrmion-like Bubbles in Perovskite Oxide Thin Films

### 6.1 Basic Properties of SrRuO$_3$ and SrIrO$_3$

Bulk SrRuO$_3$ (SRO) perovskite has a ferromagnetic Fermi liquid ground state by minority-spin double-exchange (Fig. 10a) between the electrons in the degenerate 4d $t_{2g}$ bands, with $T_C$ = 150 K[118], and pseudocubic lattice parameter ($a_{pc}$) of ~3.93 Å. Due to the high crystal field splitting $\Delta_{crys}$~2.62 eV[119] between 4d $e_g$ and $t_{2g}$ bands of Ru$^{4+}$, $e_g$ band is empty, and the $t_{2g}^4$ band has 2 unpaired spin and 2 paired spin, contributing to S=1. With SOC between S=1 and L=1, a final magnetic moment of ~1.6 $\mu_B$/Ru$^{4+}$ is resulted[120], together with large PMA for SRO thin film that is stronger than the in-plane shape anisotropy. Each $t_{2g}$ ($n^{th}$) band with non-trivial topology in k-space also carries non-zero a Chern number[121] $Z_n = \frac{1}{2\pi}\int_{BZ} \Omega_{n,z}(k)\, d^2k$, therefore endowing large intrinsic AHE[122, 123]. To be specific, the AHE loop is negative-sign due to the minority-spin double-exchange, with $P_s$ value of around -9.5% as measured by point-contact Andreev reflection[124]. For monoclinic (m-) phase of SRO thick films, optimized growth process is usually via Pulsed laser deposition (PLD) at standard 100 mTorr (13 Pa) O$_2$ pressure, 600-680 °C temperature and cooled down at >200 Torr of oxygen ambience to remove oxygen vacancy ($V_O$). Hence, its properties could be bulk-like, with octahedral tilt described by Glazer notation a$^-$b$^+$c$^-$. In the ultrathin regime of SRO along (001) orientation, lifting of degeneracy among $t_{2g}$ bands and preferential occupation of the energetically lower $d_{xy}$ orbital will promote antiferromagnetic superexchange among Ru$^{4+}$ species (Fig. 10b)[125, 126]. Tight-binding calculations with the SRO $t_{2g}$ band structure have predicted the intrinsic AHE sign-reversal due to reduction of Ru moment[121]. Besides, reducing the growth pressure and introducing $V_O$ will expand the c-axis[127], suppress the octahedral tilt and form the tetragonal (t-) phase SRO with a Glazer notation of a$^0$a$^0$c$^-$. Correspondingly, positive-sign AHE loop will dominate up to even large SRO thickness, which can be associated with extrinsic AHE mechanisms by a careful scaling analysis but is not yet done.

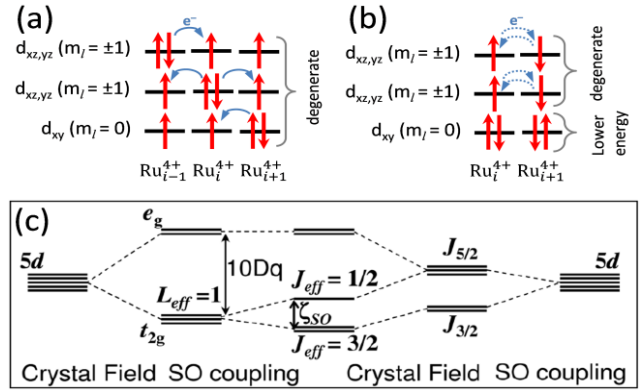

**Figure 10.** (a) Ferromagnetic minority-spin double-exchange and (b) antiferromagnetic superexchange in SRO. (c) Effect of crystal field $\Delta_{crys}$ = 10Dq and SOC splittings on electronic band structures for 5d B-site perovskites. *(c) is reprinted (adapted) with permission from Ref. [128]. Copyright (2008) American Physical Society.*

For bulk SrIrO$_3$ (SIO) with $a_{pc}$~3.94 Å, the Ir$^{4+}$ (5d $t_{2g}^5$ $e_g^0$) in perovskite crystal field also has empty $e_g$ due to the high $\Delta_{crys}$ of ~3 eV. Strong SOC ~0.5 eV splits the $t_{2g}$ band into the upper the J$_{eff}$ = 3/2 and lower J$_{eff}$ = 1/2 bands which are entangled states of the $t_{2g}$ orbitals (Fig. 10c). The 3 pairs of undistorted $t_{2g}$ wavefunctions $|J, \pm m_J\rangle$ are thus ($\sigma_\pm$ denotes spin up/down):

$$\left|\frac{1}{2}, \pm\frac{1}{2}\right\rangle = \sqrt{\frac{1}{3}}\left(|d_{yz}, \sigma_{\mp}\rangle \pm i|d_{xz}, \sigma_{\mp}\rangle \pm |d_{xy}, \sigma_{\pm}\rangle\right)$$

$$\left|\frac{3}{2}, \pm\frac{3}{2}\right\rangle = \sqrt{\frac{1}{2}}\left(|d_{yz}, \sigma_{\pm}\rangle \pm i|d_{xz}, \sigma_{\pm}\rangle\right)$$

$$\left|\frac{3}{2}, \pm\frac{1}{2}\right\rangle = \sqrt{\frac{1}{6}}\left(|d_{yz}, \sigma_{\mp}\rangle \pm i|d_{xz}, \sigma_{\mp}\rangle - 2|d_{xy}, \sigma_{\pm}\rangle\right)$$

The 5 electrons $Ir^{4+}$ thus fully fill the $J_{eff}$ = 3/2 band by 4 electrons, and half-fill the $J_{eff}$ = 1/2 Kramers doublet by 1 electron[128, 129], and can be equivalently represented by a 1-hole state with the $\left|\frac{1}{2}, \pm\frac{1}{2}\right\rangle$ wavefunction. This way, bulk $SrIrO_3$ perovskite is a paramagnetic semimetal with no magnetic long range order at ground state[130, 131]. The $SrIrO_3$ has a Dirac-like (massless) electron dispersion but heavy hole dispersion coinciding at Γ-point. The half-filled $J_{eff}$ = 1/2 band of $Ir^{4+}$ is sensitive to bandwidth (W) tuning, and large Mott Hubbard correlation factor U/W would open a Mott gap in the $J_{eff}$ = 1/2 band into upper and lower bands, akin to $Sr_2IrO_4$. If a tetragonal distortion is present, the degeneracy among the $d_{xz}$, $d_{yz}$ and $d_{xy}$ orbitals in superposition will be lifted, according to the usual crystal-field (repulsion between B-site d-orbital and anion) concept[132]. Particularly, $SrIrO_3$ and $CaIrO_3$ ultrathin films with suppressed inter-plane coupling may easily transform into weakly ordered quasi two-dimensional canted antiferromagnets[133-135]. Such magnetic phase transition is of higher-order, such that a clear $T_N$ cannot be clearly identified according to Mermin-Wagner theorem[134, 135]. Besides, according to Zhong's calculation, $Ir^{4+}$ is a potent electron donor while interfacing with another perovskite B-site cation species[136]. When such electron transfer occurs, $Ir^{4+}$ becomes hole-doped and a weakly ferromagnetic spin-liquid phase may arise assisted by proximity effect with another magnetic material that co-forms the interface[137].

*6.2 Analyses and Tuning of Hall Effects in SRO/SIO bilayers*

J. Matsuno was the first who paid attention to the THE signal of $SrIrO_3$ (2uc)/$SrRuO_3$ (m=4-7uc) grown epitaxially on $SrTiO_3$ (001), with tuning of $SrRuO_3$ thickness (m-uc)[138]. The expected trends with reducing SRO thickness are carrier localization, reducing $T_{Curie}$, and an AHE sign-reversal from negative (bulk-like) to positive which scales with diminishing magnetization. Appreciable humps resembling THE emerged only within 4-5 uc SRO, from 5 K to 80 K (Fig. 11a). Decomposition of the total Hall Effect into OHE, AHE and THE components following Eqn. 7 was done, by fitting OHE to high field background, and AHE to follow the $H_c$ and square loop shape extracted from magnetometry. THE is thus showing anti-symmetric hump shape. Using values for $R_o$ from high-field Hall slope, literature-reported $P_s$ = -9.5% and Q=1, the Skyrmion diameter was estimated to be around 5-15 nm from THE. The diminishing THE trend with increasing SRO thickness is consistent to their micromagnetic simulation using the Hamiltonian at Eqn. 2, by considering the weakening effective DMI by $D_{eff}$=D/m. The authors provided real-space imaging by low-temperature MFM to support the presence of Skyrmion-like bubbles near $H_c$.

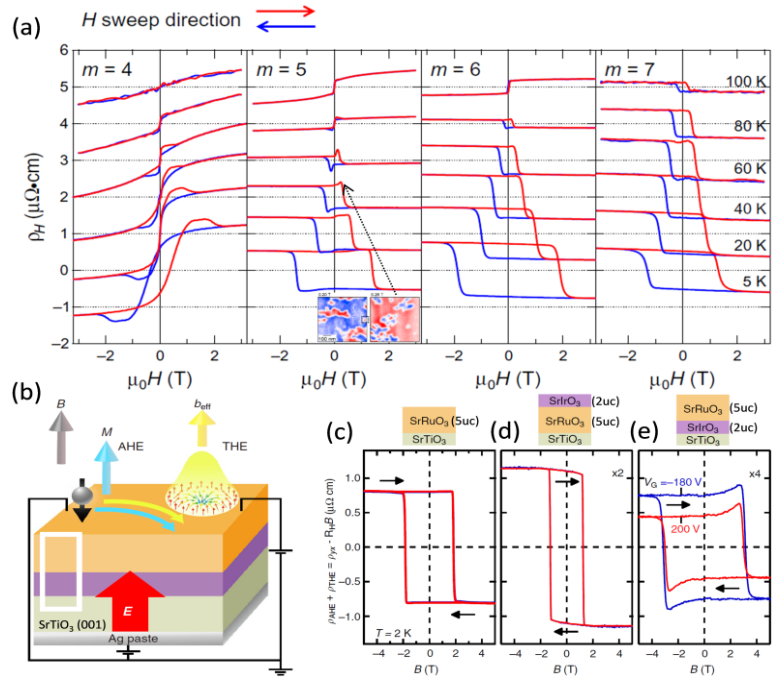

**Figure 11.** (a) Total Hall Effect loops measured from a SIO(2uc)/SRO(m-uc) bilayer by Matsuno et al., showing emergence of hump-shape THE signal in the 5-80 K temperature range for 4uc and 5uc of SRO. The MFM data in the inset shows the appearance of bubbles at THE emergence. (b) The back gating method used by Ohuchi et al. with silver-paste as bottom electrode and SRO as top electrode. (c)-(d) Gating-tuning effect is absent in the structures with SRO adjacent to STO(001) where no THE humps could be discerned, but is successful in SRO/SIO//STO(001) (e) for both AHE and THE. *Reprinted (adapted) with permission from Ref.* [138, 139]. *Copyright (2016) Science AAAS and (2018) Nature Publishing Group.*

Subsequently, electric field tuning of THE signals became a highly-pursued aspect. Y. Ohuchi et al. attempted back-gating via the STO(001) substrate (Fig. 11b), and STO is suitable due to its moderately high dielectric constant. We would expect the electric field tuning is effective on material with low carrier density. Hence, Ohuchi's result showed that it is effective when an ultrathin (2uc) SIO is inserted between STO and SRO (Fig. 11e), and at low temperature of 2 K where SIO has severe carrier localization; but not effective when the 5uc SRO is adjacent to STO (Fig. 11c-d). Likewise, if the SIO thickness adjacent to STO is increased, electric-field tuning effect also vanished, since it is screened by the mobile carriers in SIO. Upon switching electric field in SRO(5uc)/SIO(2uc)//STO(001), both AHE and THE signals are tuned. The author proposed an explanation where an electric field parallel to the broken interface inversion symmetry vector could modulate the Rashba-type SOC and DMI leading to Skyrmion size variation. Another insight is that collinear ferromagnetic SRO without Skyrmion-like textures (AHE only) cannot be tuned by electric field due to absence of magnetoelectric coupling.

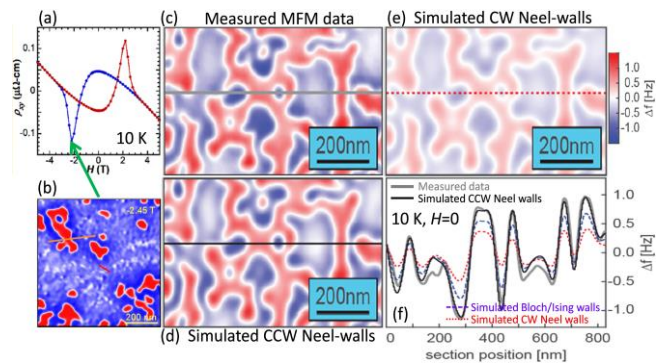

**Figure 12.** (a) Hall Effect showing strong THE hump signal, and (b) the MFM image corresponding to the THE hump emergence. (c) MFM image of random stripes at 10 K and zero-field after zero-field cooling. Simulated (d) CCW and (e) CW Néel walls with a fixed value 1.288 $\mu_B$/Ru of magnetic moment. Simulated image for Bloch/Ising walls is not shown here but is available in the original paper. (f) A summary of the MFM contrasts at the cross-sectional lines of the respective images. *Reprinted (adapted) with permission from Ref.* [140]. *Copyright (2019) American Chemical Society.*

A more detailed MFM study was provided by K. Meng et al., whose SIO(2uc)/SRO(10uc)//STO(001) bilayers were grown by off-axis sputtering[140]. One would notice that although the SRO thickness used was large, the $T_{Curie}$ is low ~110 K and all Hall loops showed AHE>0 with strong hump features, indicating possible low $\mu_B$/Ru moment (within the dead-layer regime). AHE<0 was recovered upon increasing the SRO thickness to 25uc, which is far thicker compared to typical PLD-grown films. Nevertheless, it is known that a low $\mu_B$/Ru may be advantageous for widening the phase space of THE and Skyrmions. They noticed that Hall Effect loops have very different $H_c$ compared to *M-H* loops, hence even at temperatures where Hall Effect loops do not show humps, a subtraction of expected AHE referenced to *M-H* loops would still produce strong THE humps. With MFM imaging and field sweep, maze-like domains at $H$=0 T are transformed into collinear ferromagnetic at $H$=+2.75 T then to bubbles at $H$=-2.45 T (Fig. 12a-b). Lastly, the authors point out a good way to distinguish the underlying mechanisms for stabilizing bubbles. The net magnetic charge of $\rho_{mag} = -\nabla \cdot \mathbf{M} \neq 0$ is found at zero-field which is unique to Néel-walls but not Bloch walls or Ising (collinear) walls. By a careful calibration of the MFM tip contrast[141] and performing simulations[142] using a reasonable value of 1.288 $\mu_B$/Ru, the authors concluded that CCW Néel walls nicely matched the experimental MFM data; but the other types of domain walls would produce too weak contrast (Fig. 12c-f), or require unreasonably large $\mu_B$/Ru to match the experimental data.

*6.3 Tuning of Hall Effects in SRO Single Layers*

Next, single layer SRO without SIO was also discovered to host hump-shape THE signal in several groups. Q. Qin et al. employed a t-phase SRO to demonstrate the electric field tuning of THE signal by ionic liquid gating (Fig. 13a-b)[143]. They found that electron accumulation (depletion) by applying a positive (negative) voltage to t-SRO will produce negative (positive) AHE with enhanced (reduced) THE signals, respectively. Focusing on AHE sign-reversal only, Qin's result is somewhat opposite to Ohuchi's result, where a positive gate voltage has led to sign-reversal from negative to positive AHE. Q. Qin et al. also noted that the single-layer t-phase SRO showed wider THE temperature range and electric field tunability, as compared to m-phase SRO.

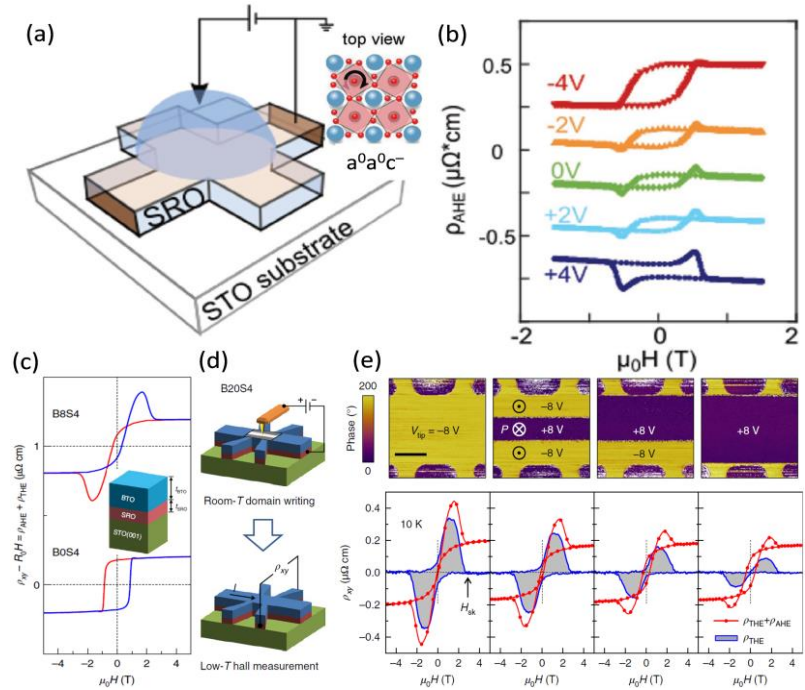

**Figure 13.** (a) A schematic showing the ionic liquid gating setup for the single layer t-phase SRO done by Q. Qin et al., and its results are shown in (b). Inset of (a) shows the top view of the t-phase SRO. (c)-(e) The BTO/SRO field-effect modulation attempted by Wang et al., with the sample schematic shown in the inset of (c). (c) Absence of BTO suppresses the hump-shape THE signal. (d) Schematics showing the sequence of BTO's ferroelectric domain writing and electrical transport measurement. (e) From left to right: increasing the $P_{DOWN}$ domain fraction (top panel) diminishes both the AHE and THE signals (bottom panel). *Reprinted (adapted) with permission from Ref.* [143, 144]. *Copyright (2019) American Chemical Society and Nature Publishing Group.*

L. Wang et al. also fabricated BaTiO$_3$ (BTO) thin film on a single-layer SRO Hall bar to demonstrate the non-volatile switching of THE[144]. The ferroelectric polarization (*P*) of BTO is switched by a piezoresponse force microscopy (PFM) at room temperature, followed by cooling down the sample to low temperature for Hall Effect or MFM measurement (Fig. 13d). Firstly, their result showed that there is an extra BTO/SRO interface contribution to the observed THE signal even without polarization switching (Fig. 13c). Secondly, area density of magnetic bubbles evolution with magnetic field can be quantified by MFM imaging, particularly reaching the densest at THE peaks. Using the same approach in Eqn. 7, the Skyrmion diameter was estimated to be ~55 nm. Thirdly, by increasing the ratio of $P_{DOWN}$ to $P_{UP}$ domains within the Hall bar channel, both AHE and THE can be reduced (Fig. 13e). Using a Density-Functional Theory (DFT) calculation supported by high-resolution scanning transmission electron microscopy (HR-STEM), the author claimed that the Ru-O bond angle and length at the BTO/SRO interface is slightly changed by the $P_{BTO}$ switching. In other words, interface SRO becomes polar where the Ru$^{4+}$ is displaced away from oxygen cage centre in the same direction as Ti$^{4+}$ displacement. In a bent Ru$_1$-O-Ru$_2$ bond, we can understand that the DMI coefficient vector $\boldsymbol{D}_{12} \propto (\boldsymbol{r}_1 \times \boldsymbol{r}_2)$, where $\boldsymbol{r}_{1,2}$ are the Ru-O bond vectors and are perpendicular to $\boldsymbol{D}_{12}$. In addition, since the increased bond bending only happens at interface, the DMI does not cancel out as in centrosymmetric bulk perovskites. Hence, $P_{UP}$ ($P_{DOWN}$) will increase (decrease) the bond bending and magnitude of $\boldsymbol{D}$, thus tuning the Skyrmion density and size. Besides, since Ru-O-Ru bond bending cannot be inverted due to the intrinsic octahedral tilts/rotation, the $\boldsymbol{D}$ sign-reversal does not occur. Conversely, if $\boldsymbol{D}$ were to invert sign, we could expect changes only in the Skyrmion's helicity $Q_h$ but the topological charge $Q$ and thus $\rho_{THE}$ should maintain the same sign at a fixed magnetic field. We believe that this electric field gating approach via ferroelectric switching is closer to true atomic-scale DMI tuning compared to the earlier back-gating and ionic liquid gating approaches.

Almost concurrently, W. Wang et al. provided an insight about THE from a quite different phase space, which is near and above $T_{Curie}$ in single layer SRO and V-doped Sb$_2$Te$_3$ topological insulator (TI)[83]. This corresponds to the random chirality fluctuation regime similar to the earlier La$_{1-x}$Ca$_x$MnO$_3$ cases[82] as discussed in the previous section (Fig. 14a-b). Judging from the PLD growth parameter for SRO of 650 °C and 0.25 mbar, the structural phase should m-phase. Due to an additional capping layer of 2uc STO on the 6uc SRO (Fig. 14c), out-of-plane $K_u$ is enhanced, and no THE hump signal could be discerned at low temperatures; but the THE signals spotted around $T_{Curie}$~116 K is non-hysteretic since the $H_c$ at that regime is vanishingly small (Fig. 14e-f). The THE antisymmetric humps can fitted well with $\rho_{THE}(H) \sim \frac{M(H)}{1+(H/D_{eff})^2}$ where $M(H)$ should take the form of Langevin function, and $D_{eff}$ of ~0.2 meV was extracted. With similar Hall data processing techniques, the THE signal is maximum at $T_{Curie}$ but decays fast around it. Similar result was also obtained in the V:Sb$_2$Te$_3$//STO(111) by tuning temperature and electric field gating voltage, although with an opposite-sign $\rho_{THE}$. The author supported these trends with Monte-Carlo simulation for chirality fluctuation by: $\exp\left(-\frac{i\phi_{ijk}}{2}\right) = \frac{1+\hat{m}_i \cdot \hat{m}_j + \hat{m}_j \cdot \hat{m}_k + \hat{m}_k \cdot \hat{m}_i + i(\hat{m}_i \times \hat{m}_j) \cdot \hat{m}_k}{\sqrt{2(1+\hat{m}_i \cdot \hat{m}_j)(1+\hat{m}_j \cdot \hat{m}_k)(1+\hat{m}_k \cdot \hat{m}_i)}}$ and $Q = \frac{1}{4\pi}\sum_{ijk}\phi_{ijk}$ where i,j,k are indices for neighbouring moments in a triangle (Fig. 14d).

Most recently, hydrogen-doping was injected into SRO thin film by ionic liquid gating, as performed by Pu Yu et. al, causing structural lattice expansion and weakened ferromagnetism[145]. Surprisingly, THE signals also emerged in 2-100 K by an optimal gate voltage.

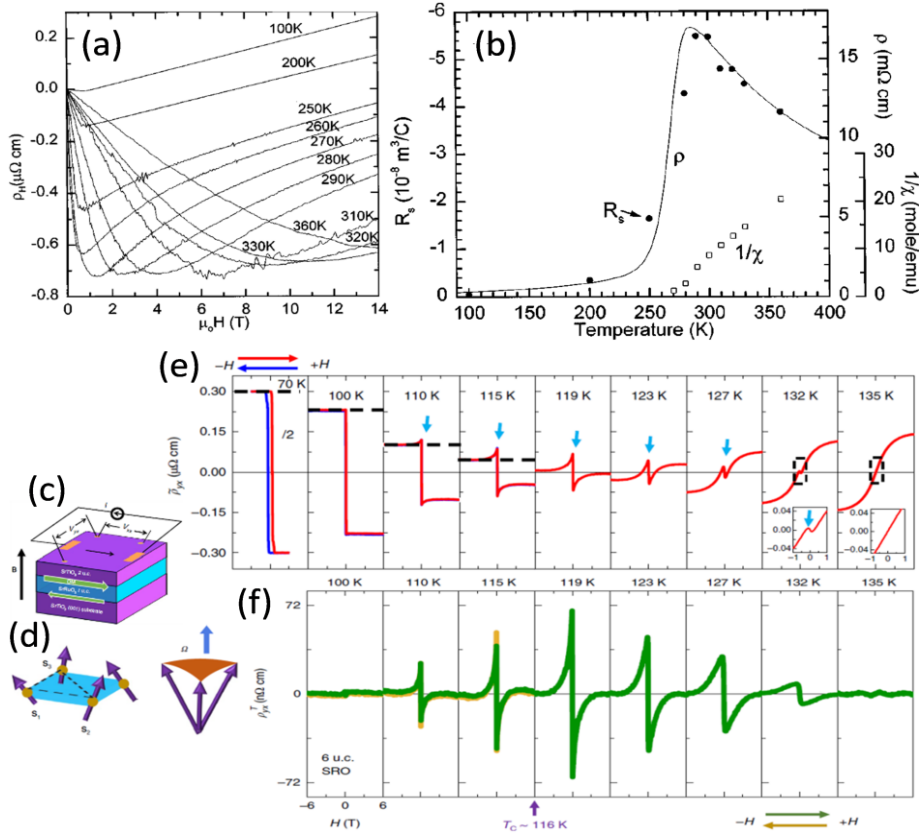

**Figure 14.** For $La_{1-x}Ca_xMnO_3$ studied much earlier by Chun et al., the Hall Effect data (a) showed weak humps, and (b) the Anomalous Hall Coefficient $R_s$ shows matching temperature dependence with linear resistivity ($\rho_{xx}$), and $1/\chi$ from DC magnetometry shows paramagnetic behaviour after $T_{Curie}$. (c)-(f) Spin Chirality fluctuation behaviour similar to that of $La_{1-x}Ca_xMnO_3$ was recently obtained by W. Wang et al., using the SRO heterostructure shown at (c). (d) The explanation of Berry phase from solid angle subtended by the moment triads. Hall data (AHE+THE in (e) and pure THE in (f)) shows strongest signal of non-hysteretic THE humps around $T_{Curie}$. Reprinted (adapted) with permission from Ref. [81, 83]. Copyright (1999) American Physical Society and (2019) Nature Publishing Group.

*6.4 Emergence of THE in Ce-doped $CaMnO_3$ with strong Electron Correlations*

On the other hand, L. Vistoli et al. found strong THE signals in perovskite $Ce_xCa_{1-x}MnO_3$ with $Ce^{4+}$ doping of 0<x<5%[146]. Such doping creates a weakly ferromagnetic metallic phase with $T_{Curie}$~100 K (Fig. 15a-b), which has been a lattice-matched bottom electrode material widely used for earlier $BiFeO_3$-based ferroelectric tunnelling junctions (FTJ)[147]. Particularly, for x=4%, THE signal spanned across a wide temperature range of 15-75 K (Fig. 15c). Their low-temperature MFM images support the presence of magnetic bubbles and have almost the same quality as what attempted by J. Matsuno and K. Meng et.al. mentioned above. Since there is no obvious sharp interface with a heavy metal providing SOC and Rashba-type DMI, nor there is a non-centrosymmetric crystal structure, the author used the earlier approach in dipolar/demagnetization-stabilized bubbles (as discussed in section 2). Namely, the bubbles diameter was estimated by $32\frac{\sqrt{J_{ex}K_u}}{\mu_0 M_S^2}$ to be ~250 nm, agreeing with MFM measurement, where the $J_{ex}$, $K_u$ and $M_S$ were calculated from $T_N$, magnetometry loops and moment-canting angle respectively. A particular strong selling point of this paper is on the superfast scaling of THE signal with carrier density (*n*) (Fig. 15d), which was found to deviate significantly from the simple Bruno's model of strong $\hat{\sigma}$-$\hat{m}$ (adiabatic) interaction regime (Eqn. 7). Particularly at low Ce-doping, $\rho_{THE}$ is enhanced by $\propto n^{-8/3}$ together with enhanced electron effective mass $m_e^*$ due to strong electron correlation, consistent to the "weak coupling, localized magnetic texture" regime as theorized by Nakazawa et al.[111] (as well as suppl. info. of Ref. [146]).

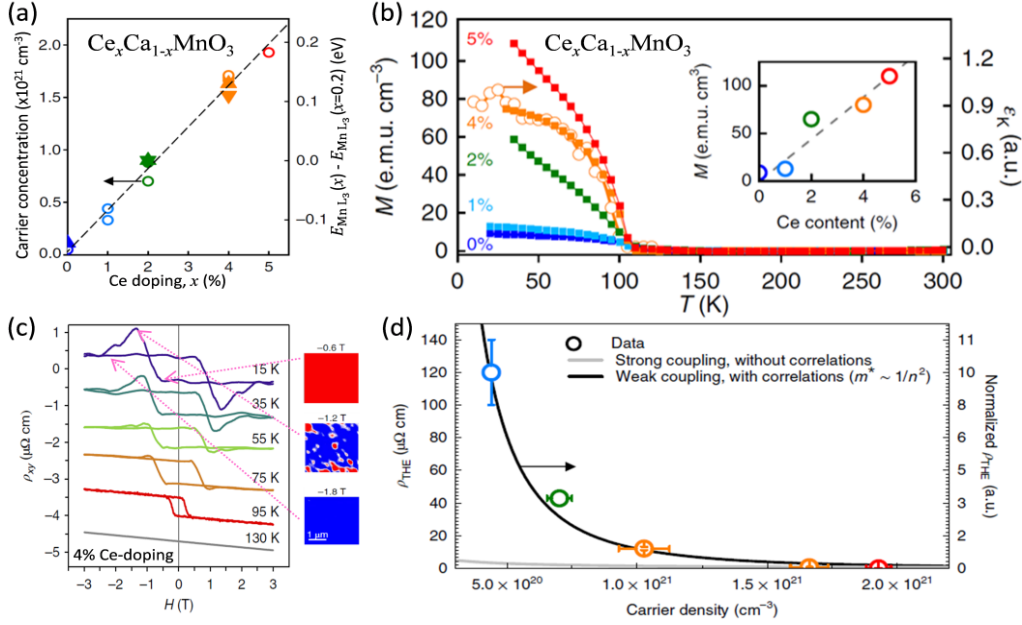

**Figure 15.** Increasing (a) carrier concentration and (b) saturated magnetization with Ce-doping in CaMnO$_3$ below 5%. (c) Hall data of 4% Ce-doping showing strong THE signals, supported by MFM images (inset). After saturating at +3T, MFM images were obtained in sequence from -0.6 T, -1.2 T to -3T. (d) Exponential dependence of THE signal with carrier density plotted together with its fitting models. *Reprinted (adapted) with permission from Ref. [146]. Copyright (2019) Nature Publishing Group.*

*6.5 The Alternative "Bi-AHE" Interpretation*

Several other groups have provided an alternative interpretation for the hump signals to be artefacts of partial cancellation between two AHE loops, which can be modelled by two pairs of overlapping Langevin functions $\mathcal{L}_{12}(H)$ with opposite sign of coefficients and different $H_{c1,2}$ (Fig. 16a). This way, at the hump emergence, only trivial Ising-like (collinear) domain walls exist instead of non-coplanar textures. In particular, Kan[148] and Groenendijk[121] et al. performed such Bi-AHE analyses on the hump-shape Hall features of the SRO-based heterostructures. Using a DFT calculation combined with a tight-binding model for Wannier function interpolation, the possible AHE sign-reversal at interfaces between Ru/Ir, Ru/Ru and Ru/Ti was calculated, as shown in Fig. 16b. Besides, they performed minor Hall loop measurement (Fig. 16c-d) and revealed that upon reaching the hump peaks and without reaching high-field saturation, the peak $\rho_{xy}$ value can be retained by sweeping back to zero-field. We opine that while such behaviour looks very supportive to the picture of partial cancellation of two AHE loops, it could also be a hysteretic behaviour of Skyrmion-like bubbles, i.e. Skyrmions can stabilize at the phase space of multi-*q* helical/cycloidal stripes.

This debate points to the importance of demanding a reliable magnetic imaging technique. In parallel to the "Bi-AHE" interpretation, L. Wang et al. performed another MFM experiment on single layer SRO film comparing fast and slow PLD growth rate[149]. It was found that the inhomogeneity of number of SRO unit cells along atomic terraces in step-flow growth mode (Fig. 16g) is the source of two opposite-sign AHE loops, and correspondingly their MFM data showed trivial domains (Fig. 16f). In particular, when the growth rate is slowed, uniform and evenly distributed 4uc and 5uc regions can be found along a single terrace, while the 4uc and 5uc regions contribute positive- and negative-sign AHE respectively (Fig. 16h-i). Hence the net AHE resulted was completely cancelled at saturation but huge humps emerged near $H_c$, since the $H_c$ differences between the two AHE contributions causes incomplete cancellation (Fig. 16e), when the two different magnetic domains are antiparallel.

As a concluding remark for this sub-section, notably, obtaining high-resolution MFM images on Skyrmions is very challenging, and deducing the actual Skyrmion type (Néel/Bloch) remained unclear or indirect. To probe deeper, the diamond NVM capable of single-spin sensitivity in constructing the smoothly varying magnetic texture, as achieved by Dovzhenko et al.[58], would be very appreciable. Alternatively, if oxide thin films can be detached from the thick substrates by a suitable epitaxial-lift off technique[150], L-TEM would be again useful, either with or without slight tilting of electron beam vector ***k*** away from sample's surface normal for imaging Néel-[56] or Bloch-type Skyrmions, respectively. The possibility of measuring the interface DMI strength using Brillouin Light Scattering (BLS)[151] in oxide heterostructures is also almost ruled out since the reflective surface as required in BLS is absent, unlike true metals. Furthermore, in the work on electric-field tuning of Hall Effect signals by Y. Ohuchi, Q. Qin, and L. Wang[139, 143, 144], the possibility of $V_O$ movement in SRO that contributes AHE tuning has not been ruled out. Nevertheless, tuning the surface bond-bending angle of a non-multiferroic ferromagnetic material via polarization switching of an adjacent ferroelectric film is certainly a good strategy of modulating the interface-DMI strength. In short, we believe the THE phenomenon in oxide thin films remains as an emergent field at infancy stage, with more perplexed Physics to be excavated.

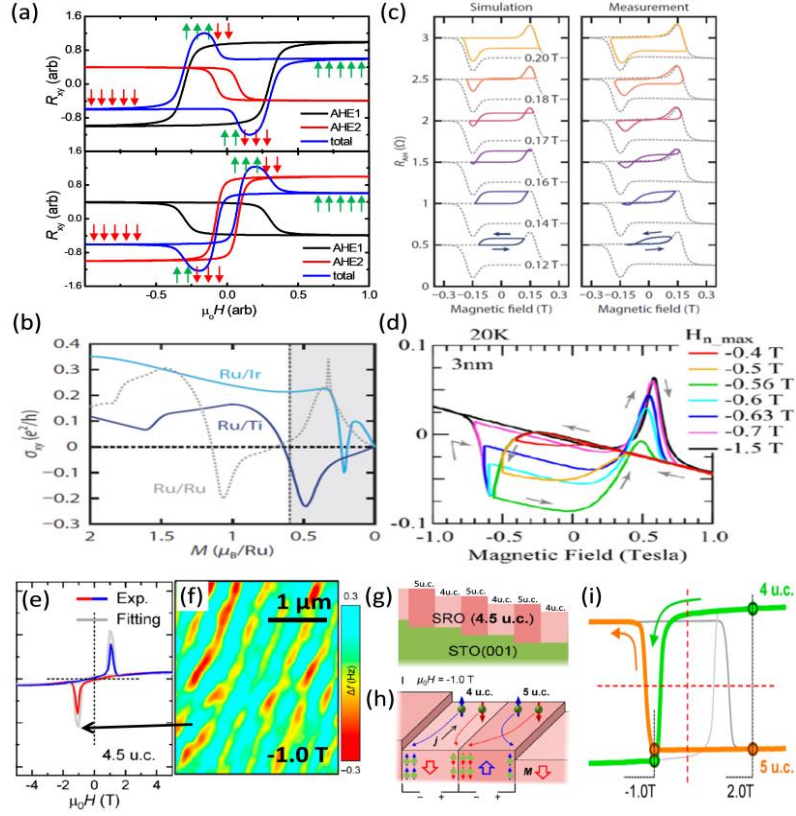

**Figure 16.** (a) Top and bottom: two possible ways of humps emergence in Hall measurement and can be reproduced by a summation of AHE1 and AHE2 loops of opposite-sign and different $H_c$. (b) Calculated variations in Anomalous Hall conductivity with Ru-moment at different interfaces. (c) and (d) Minor loop measurements performed by the two authors mentioned. The $H_{n\_max}$ in (d) means maximum negative field. (e) Complete AHE cancellation for 4.5 u.c. SRO, with MFM data (f) at -1.0 T showing uniform domains instead of bubbles. (g) A sketch showing the uniformly alternating 4uc and 5uc regions distributed on atomic terraces of the average 4.5 u.c. SRO sample. (h) and (i) The 5uc region is understood to have larger $H_c$, hence only the 4uc region is switched down at -1.0 T. The reason of this thickness-driven AHE sign-reversal can be inferred from section 6.1 where a lower moment in 4uc SRO will produce a positive AHE loop. *Reprinted (adapted) with permission from Ref.* [121, 148, 149]. *Copyright 2019) American Physical Society and (2019) American Chemical Society.*

*6.6 Skyrmions at Ferrimagnetic Oxide/heavy metal bilayers*

Here, we review about iron-based ferrimagnets bringing together two valuable properties – having vanishingly small $M_S$ near its compensation temperatures as well as high (>300 K) magnetic ordering temperatures. Assisted by an adjacent ultrathin heavy metal layer, such properties are good candidates for stabilizing ultra-small Skyrmions (<10 nm diameters) at room temperature[152], which are also highly mobile[153] with reduced Skyrmion Hall Effect[45]. These properties are superior compared to the THE signals observed in the Ru-, Ir- and Mn-based oxide perovskites discussed in the earlier sub-sections.

Büttner et al. developed a full micromagnetic stray-field model that bridges the gap between the domain-wall model[67, 68] (Eqn. 5) and the effective anisotropy model[10] (Eqn. 3). They noticed that in an $M_S$-versus-$K_u$ phase diagram (Fig. 17a-b), conventional metallic ferromagnets or their multilayers are located in the large $M_S$ but small to moderate $K_u$ regime. Hence, they do not satisfy the condition of small Skyrmions stabilized at room temperature and zero/weak magnetic field[152], but produce the stripe phase instead. To possibly reach the small Skyrmion (~10 nm) regime at room temperature, the low $M_S$ and high $K_u$ conditions are required, and ferrimagnets are better candidates for this purpose.

Besides, L. Caretta et al. demonstrated the benefit of using ferrimagnetic $Gd_{44}Co_{56}$ to achieve ultrafast domain-walls' or Skyrmions' speed[153] – up to >1 km/s. Ferrimagnets typically have two unique temperatures – the magnetization compensation temperature ($T_M$) and the angular momentum compensation temperature ($T_A$) (Fig. 17c). At $T_M$, $M_{1,2}$ of the two antiferromagnetically coupled sub-lattices cancel, i.e. $M(T) = M_1+M_2 = 0$, resulting in divergence of $H_c$. Whereas at $T_A$, their spin angular momentum $S(T) = M_1/\gamma_1 + M_2/\gamma_2 = 0$, resulting in divergence of domain wall velocity. The difference between the gyromagnetic coefficients of the two sub-lattices $\gamma_{1,2} = g_{1,2}\mu_B/\hbar$ implies $T_A$ and $T_M$ are separated but near. From the SHE-driven DW velocity, the effective field of damping-like SOT, $H_{eff} = \frac{\pi\hbar\theta_{SH}^{eff}j_{HM}}{4e\mu_0 M_s t}$ should diverge at $T_M$ (near $T_A$). It also becomes very efficient when $S(T)$ vanishes at $T_A$, resulting in the huge increase in SHE-driven DW velocity, $v_{FI} = \frac{\pi}{2}\frac{Dj_{HM}}{\sqrt{(S(T)j_{HM})^2+(S_o j_o)^2}}$ for ferrimagnets, where $j_o = \frac{2etD}{\hbar\theta_{SH}^{eff}\Delta}$ is related to the DMI, effective spin-hall angle ($\theta_{SH}^{eff}$), and DW width ($\Delta$). Hence at $T_A$, the ferrimagnetic quasiparticles (such as DW or Skyrmions) would behave like Newtonian particles without velocity saturation even at large current densities (Fig. 17d-e). Conversely, the DW velocity of a conventional ferromagnet, $v_{FM} = \frac{\pi\gamma}{2M_s}\frac{Dj_{HM}}{\sqrt{j_{HM}^2+(\alpha j_o)^2}}$ is limited by the large $M_S$ and would saturate when $j_{HM}\gg\alpha j_o$, where $\alpha$ is the Gilbert damping constant. Insulating ferrimagnetic oxides also bring the advantage of having lower $\alpha$ (~$10^{-4}$) compared to metallic ferromagnets (~$10^{-2}$), implying that lower $j_{HM}$ is needed for driving domain-walls or Skyrmions.

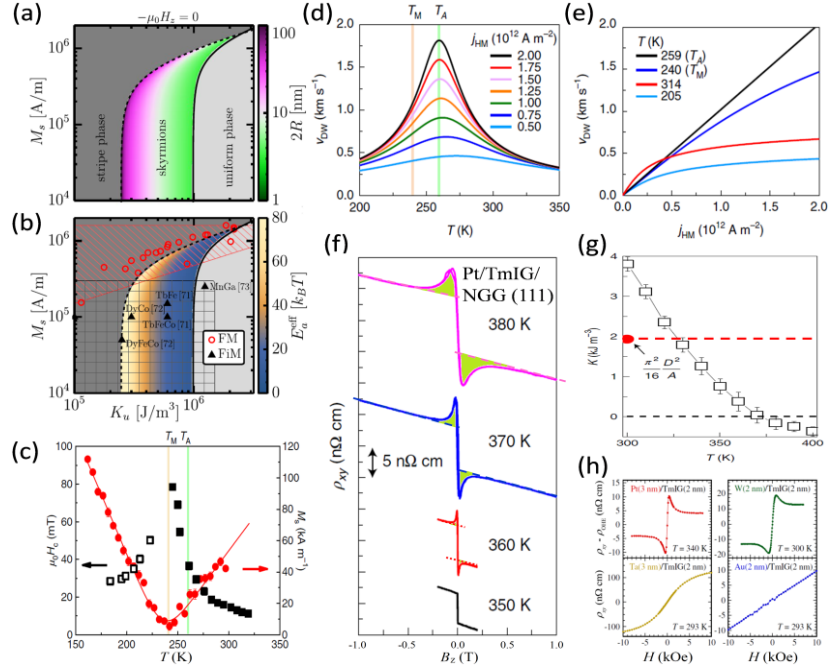

**Figure 17:** (a) $M_S$-$K_u$ phase diagram from Buttner's full stray-field model built with near-realistic parameters with colour bar showing Skyrmion diameters. (b) The same stray-field model overlapped with experimental data (circles and triangles), with colour bar showing their stabilizing energy. (c) Temperature-dependent $H_c$ and $M_S$ variations across $T_M$ and $T_A$ for $Co_{44}Gd_{56}$. The calculated SOT-driven domain-wall velocity in Pt/$Co_{44}Gd_{56}$ varying with (d) temperature and (e) current density agrees well with experimental data. (f) Hall Effect of Pt/TmIG/NGG(111) with THE signals shaded in green, and data are shifted vertically for clarity. Its anisotropy is measured by magnetometry (g) with the dotted lines marking the threshold below which Skyrmions are expected to emerge. (h) Hall Effect from various heavy metal/TmIG (111) bilayers. *Reprinted (adapted) with permission from Ref. [152-155]. Copyright (2018-2019) Nature Publishing Group, and (2020) American Physical Society.*

Q. Shao et al. first observed THE signal in a Pt/$Tm_3Fe_5O_{12}$ (TmIG)//$Nd_3Ga_5O_{12}$(111)[154] (Fig. 17f-g). The THE signal appears above 300 K where $K_u < \frac{\pi^2 D^2}{16A}$, justifying the presence of DMI-stabilized Skyrmions. The THE signal is also distinct from those observed in $SrRuO_3$ systems reported earlier, since it is not hysteretic – Hall peak features appear at small fields when the field is swept towards both positive and negative directions. Such non-hysteretic behavior could be understood as the result of low coercive field $\mu_0 H_c < 10$ mT of typical ferrimagnetic garnets. Furthermore, such non-hysteretic THE signal is also convincing since it avoids the debated interpretation arisen from partial cancellation of two oppositely-signed AHE loops. Fengyuan Yang's group in Ohio University subsequently coined this effect as "Spin-Hall Topological Hall Effect (SH-THE)", describing that the SHE in ultrathin Pt causes (vertical) spin accumulation at the interface, which is able to detect the presence of topological magnetic textures at the Pt/TmIG interface[156]. Comparing between the strain effect from $Gd_3Ga_5O_{12}$ (GGG) and substituted GGG (s-GGG) substrates on TmIG film, the larger compressive strain on TmIG from s-GGG (but not GGG) is realized to induce PMA and hence the SH-THE signal. It is known that the Spin-Hall conductivity flips sign across the 5d heavy metals from Ta to Au[157]. Aidan et al. further replaced ultrathin Pt with Ta, W and Au on TmIG films[155] (Fig. 17h). While W showed similar behaviour as Pt, Ta was unable to create Skyrmions in TmIG due to its weak interfacial DMI contribution. Whereas the Au case, a strong DMI exists to create Skyrmions but its spin accumulation and detection of SH-THE are ineffective due to the long spin diffusion length in Au. Lastly, the 4f electrons in the rare-earth cations of ferrimagnetic garnets are found to contribute no obvious role in stabilizing SH-THE[158].

## 7. Future Outlook

In this final section, we briefly direct the readers to some interesting related work. Note that there is a slight contradiction from the design of Skyrmion racetrack memory in section 1. Criterion (1) for Skyrmion nucleation by dissipation-less electric field requires an insulator with low carrier density to avoid screening of the field effect; yet criterion (2) for current-driving Skyrmions along the racetrack requires a metallic body. Such contradiction would raise the question whether an interface Skyrmion residing at an insulator-metallic material interface would be useful, i.e.: the insulating film provides the necessary magnetic properties ($M$, $K_u$) that can be tuned by electric field, while the metallic film provides the SOC and Rashba-type DMI via the sharp interface, and also functions as an electrode for the electric field gating process. Hence, a good example is a ferromagnetic interface arising from charge transfer between a heavy metal and antiferromagnetic insulator, such as the $SrMnO_3$/$SrIrO_3$ interface, which received much attention. Following Zhong's calculation based on the concept of O2p band alignment before contact[136], $Ir^{4+}$ with larger ($E_{O2p}-E_F$) gap will readily transfer electrons to $Mn^{4+}$. Nichols et al. have fabricated such superlattice (SL) and found strong negative-sign AHE but no THE hump features[159]. The similarity of AHE between SMO/SIO-SL and SRO suggest that they share similar $t_{2g}$ band structures with large amounts of Weyl nodes. Using a tight-binding Hamiltonian $\widetilde{\mathcal{H}} = -t\cos\left(\frac{\beta}{2}\right)\sum_{<ij>}\left(a_{i\uparrow}^\dagger a_{j\uparrow} + \text{h.c.}\right) + \sum_{<ij>} J_{SE}\left(\widehat{\boldsymbol{m}}_i \cdot \widehat{\boldsymbol{m}}_j\right)$, Bhowal et al.[137] explained that there exists a threshold charge-transfer fraction $x_c = \frac{2J_{SE}}{|t|}$ above which the system is ferromagnetic but below which the system is canted antiferromagnetic. Okamoto calculated that it is possible to stabilize SkL in LSMO/SIO-SL[117], and Bhowal suggested that the Rashba-type SOC and AHE in SIO/SMO-SL can be tuned by electric field on modulating the hopping integral between non-overlapping orbitals[36]. Hence such charge-transfer induced interface ferromagnetism can be a good platform for Skyrmion search, though a careful strategy is needed to avoid mutual cancellation of interface DMI. Another very promising structure is the Pt/$Tm_3Fe_5O_{12}$ interfaces where non-hysteretic THE signals were found at above room temperature[154, 156]. It's THE data is similar Fig. 9j, due to low $H_c$. For the aspect of electrical tuning THE on such structure, the electric field could be applied on the insulating ferromagnetic garnet via back-gating through the substrate with Pt as the top electrode, but has not been demonstrated yet.

Several theorists have investigated confined heavy metal oxide perovskite systems in (111)-orientation with honeycomb lattice that are potential candidates of TI[160-162]. Yet the ideal candidate such as $LaAuO_3$ may be difficult to synthesize. L. Si et al. considered the concept that confined ultrathin SRO(111) can avoid degeneracy-lifting among $t_{2g}$ orbitals, thus can avoid a phase transition to canted antiferromagnetic but remain ferromagnetic[163], but will open a small Mott gap near $E_F$. This way a Chern insulator hosting Quantum Anomalous Hall Effect (QAHE) could be stabilized in SRO(111). It would also be interesting to explore SRO/SIO//STO(111) or SMO/SIO-SL//STO(111) in search for THE, since the single-terminations of (111)-oriented perovskite are always polar and would confine moment modulation and DMI vectors towards in-plane (perpendicular to the polar direction).

Besides, a coexistence of QAHE and THE may also be formed at low temperature such as in the recent case of $Cr_x(Bi,Sb)_{2-x}Te_3$ – although the AHE part must be quantized, THE signal and possible magnetic Skyrmions can still emerge around the $H_c$ field range of magnetization reversal[164]. This way, it might be possible utilize the recent finding of Topological Magnetoelectric Effect (TME) for achieving electric field control of Skyrmion nucleation or annihilation. In the seminal work by Marguerite et al.[165], a nanoSQUID-on-tip[166] was employed to apply a localized electric field and image the circulating surface current around mirror magnetic monopoles[167] that are induced by TME. Such mirror magnetic monopole may potentially interact with the emergent magnetic field $B_{\text{eff}} = \frac{Qh}{e\pi r_{\text{sk}}^2}$ of the topological charges of non-multiferroic Skyrmions in general, therefore offering an alternative opportunity to achieve Skyrmion size tuning, creation and annihilation by electric field.

## Acknowledgements

This research is supported by the Agency for Science, Technology and Research (A*STAR) under its Advanced Manufacturing and Engineering (AME) Individual Research Grant (IRG) (A1983c0034), the National University of Singapore (NUS) Academic Research Fund (R-144-000-403-114), and the Singapore National Research Foundation (NRF) under the Competitive Research Programs (CRP Award No. NRF-CRP15-2015-01).